\documentclass[twocolumn]{aastex631}

\usepackage{amsmath}
\usepackage{graphicx}

\shorttitle{Dust Formation in SN\,2014C}
\shortauthors{Tinyanont et al.}
\begin{document}

% \title{Dust Formation and Ongoing Interactions with Asymmetric Circumstellar Medium in the Transitional SN\,2014C Observed with James Webb Space Telescope}

% \title{Dust Formation and Asymmetric Circumstellar Medium Interaction in the Transitional SN\,2014C Observed with James Webb Space Telescope Mid-Infrared Spectroscopy}

\title{Large Cold Dust Reservoir Revealed in Transitional SN~Ib~2014C by James Webb Space Telescope Mid-Infrared Spectroscopy}

\correspondingauthor{Samaporn~Tinyanont}

\author[0000-0002-1481-4676]{Samaporn~Tinyanont}
\affiliation{National Astronomical Research Institute of Thailand, 260 Moo 4, Donkaew, Maerim, Chiang Mai, 50180, Thailand}
\email{samaporn@narit.or.th}

\author[0000-0003-2238-1572]{Ori~D.~Fox}
\affiliation{Space Telescope Science Institute, Baltimore, MD 21218, USA}

\author[0000-0002-9301-5302]{Melissa~Shahbandeh}
\affiliation{Space Telescope Science Institute, Baltimore, MD 21218, USA}

\author[0000-0001-7380-3144]{Tea~Temim}
\affiliation{Department of Astrophysical Sciences, Princeton University, 4 Ivy Lane, Princeton, NJ 08544, USA}

\author[0000-0002-3742-8460]{Robert~Williams}
\affiliation{Space Telescope Science Institute, Baltimore, MD 21218, USA}

\author[0000-0002-7593-2748]{Kittipong~Wangnok}
\affiliation{National Astronomical Research Institute of Thailand, 260 Moo 4, Donkaew, Maerim, Chiang Mai, 50180, Thailand}
\affiliation{School of Physics, Institute of Science, Suranaree University of Technology, Nakhon Ratchasima, 30000, Thailand}

\author[0000-0002-4410-5387]{Armin~Rest}
\affiliation{Space Telescope Science Institute, Baltimore, MD 21218, USA}
\affiliation{Department of Physics and Astronomy, The Johns Hopkins University, Baltimore, MD 21218, USA}

\author[0000-0003-0778-0321]{Ryan~M.~Lau}
\affiliation{NSF's NOIRLab, 950 N. Cherry Avenue, Tucson, AZ 85719, USA}

\author[0000-0003-2611-7269]{Keiichi Maeda}
\affiliation{Department of Astronomy, Kyoto University, Kitashirakawa-Oiwake-cho, Sakyo-ku, Kyoto 606-8502, Japan}

\author[0000-0001-5754-4007]{Jacob~E.~Jencson}	
\affiliation{IPAC, California Institute of Technology, 1200 E. California Blvd., Pasadena, CA 91125, USA}

\author[0000-0002-4449-9152]{Katie~Auchettl}
\affiliation{Department of Astronomy and Astrophysics, University of California, Santa Cruz, CA 95064, USA}
\affiliation{School of Physics, University of Melbourne, VIC 3010, Australia}

\author[0000-0003-3460-0103]{Alexei~V.~Filippenko}
\affiliation{Department of Astronomy, University of California, Berkeley, CA 94720-3411, USA}

\author[0000-0003-2037-4619]{Conor~Larison}
\affiliation{Department of Physics \& Astronomy, Rutgers, State University of New Jersey, 136 Frelinghuysen Road, Piscataway, NJ 08854, USA}

\author[0000-0002-5221-7557]{Chris~Ashall}
\affiliation{Institute for Astronomy, University of Hawai'i at Manoa, 2680 Woodlawn Dr., Hawai'i, HI 96822, USA }

\author[0000-0001-5955-2502]{Thomas~G.~Brink}
\affiliation{Department of Astronomy, University of California, Berkeley, CA 94720-3411, USA}

\author[0000-0002-5680-4660]{Kyle~W.~Davis}
\affiliation{Department of Astronomy and Astrophysics, University of California, Santa Cruz, CA 95064, USA}

\author[0000-0003-0599-8407]{Luc~Dessart}	
\affiliation{Institut d'Astrophysique de Paris, CNRS-Sorbonne Universit\'e, 98 bis boulevard Arago, F-75014 Paris, France}

\author[0000-0002-2445-5275]{Ryan~J.~Foley}
\affiliation{Department of Astronomy and Astrophysics, University of California, Santa Cruz, CA 95064, USA}

\author[0000-0002-1296-6887]{Llu{\'i}s Galbany}
\affiliation{Institute of Space Sciences (ICE, CSIC), Campus UAB, Carrer de Can Magrans, s/n, E-08193 Barcelona, Spain}
\affiliation{Institut d'Estudis Espacials de Catalunya (IEEC), 08860 Castelldefels (Barcelona), Spain} 

\author[0000-0002-6741-983X]{Matthew~Grayling}
\affiliation{Institute of Astronomy and Kavli Institute for Cosmology, University of Cambridge, Madingley Road, Cambridge CB3 0HA, UK}

\author[0000-0001-5975-290X]{Joel~Johansson}
\affiliation{Oskar Klein Centre, Department of Physics, Stockholm University, SE-10691 Stockholm, Sweden}

\author[0000-0002-5619-4938]{Mansi~M.~Kasliwal}
\affiliation{Division of Physics, Mathematics, and Astronomy, California Institute of Technology, Pasadena, CA 91125, USA}

\author[0009-0003-8380-4003]{Zachary~G.~Lane}
\affiliation{School of Physical and Chemical Sciences — Te Kura Matū, University of Canterbury, Private Bag 4800, Christchurch 8140, New Zealand}

\author[0000-0002-2249-0595]{Natalie~LeBaron}
\affiliation{Department of Astronomy, University of California, Berkeley, CA 94720-3411, USA}

\author[0000-0002-0763-3885]{Dan~Milisavljevic}
\affiliation{Department of Physics and Astronomy, Purdue University, 525 Northwestern Ave., West Lafayette, IN 47907}

\author[0000-0003-3643-839X]{Jeonghee Rho}
\affiliation{SETI Institute, 189 Bernardo Ave., Ste. 200, Mountain View, CA 94043, USA}

\author[0000-0001-7641-5497]{Itsuki~Sakon}
\affiliation{Department of Astronomy, Graduate Schools of Science, University of Tokyo, 7-3-1 Hongo, Bunkyo-ku, Tokyo 113-0033, Japan}

\author[0000-0002-9820-679X]{Arkaprabha Sarangi}
\affiliation{Indian Institute of Astrophysics, Bengaluru 560034, India}
\affiliation{DARK, Niels Bohr Institute, Copenhagen, Denmark}

\author[0000-0003-4610-1117]{Tamás Szalai}
\affiliation{Department of Experimental Physics, Institute of Physics, University of Szeged, D{\'o}m t{\'e}r 9, 6720 Szeged, Hungary}
\affiliation{MTA-ELTE Lend\"ulet ``Momentum" Milky Way Research Group, Szent Imre H. st. 112, 9700 Szombathely, Hungary}

\author[0000-0002-5748-4558]{Kirsty~Taggart}
\affiliation{Department of Astronomy and Astrophysics, University of California, Santa Cruz, CA 95064, USA}

\author[0000-0001-9038-9950]{Schuyler~D.~Van~Dyk}
\affiliation{IPAC, California Institute of Technology, 1200 E. California Blvd., Pasadena, CA 91125, USA}

\author[0000-0001-5233-6989]{Qinan~Wang}
\affiliation{Department of Physics and Kavli Institute for Astrophysics and Space Research, Massachusetts Institute of Technology, 77 Massachusetts Avenue, Cambridge, MA 02139, USA}
\affiliation{TESS-ULTRASAT Joint Postdoctoral Fellow}

\author[0000-0002-6535-8500]{Yi~Yang}
\affiliation{Physics Department, Tsinghua University, Beijing, 100084, China}
\affiliation{Department of Astronomy, University of California, Berkeley, CA 94720-3411, USA} 

\author[0000-0002-2636-6508]{WeiKang~Zheng}
\affiliation{Department of Astronomy, University of California, Berkeley, CA 94720-3411, USA} 
\affiliation{Bengier-Winslow-Eustace Specialist in Astronomy} 

 \author[0000-0001-7473-4208]{Szanna~Zs{\'i}ros} 
 \affiliation{Department of Experimental Physics, Institute of Physics, University of Szeged, D{\'o}m t{\'e}r 9, 6720 Szeged, Hungary}

\begin{abstract}
Supernova (SN) 2014C is a rare transitional event that exploded as a hydrogen-poor, helium-rich Type Ib SN and subsequently interacted with a hydrogen-rich circumstellar medium (CSM) a few months post-explosion. 
This unique interacting object provides an opportunity to probe the mass-loss history of a stripped-envelope SN progenitor.
Using the {James Webb Space Telescope (JWST)}, we observed SN\,2014C with the Mid-Infrared Instrument Medium Resolution Spectrometer at 3477 days post-explosion (rest frame), and the Near-Infrared Spectrograph Integral Field Unit at 3568 days post-explosion, covering 1.7 to 25 $\mu$m.
The bolometric luminosity indicates that the SN is still interacting with the same CSM that was observed with the {Spitzer Space Telescope} 40--1920 days post-explosion. 
{JWST} spectra and near-contemporaneous optical and near-infrared spectra show strong [\ion{Ne}{2}]~12.831~$\mu$m, He~1.083~$\mu$m, H$\alpha$, and forbidden oxygen ([\ion{O}{1}] $\lambda$$\lambda$6300, 6364, [\ion{O}{2}] $\lambda$$\lambda$7319, 7330, and [\ion{O}{3}] $\lambda$$\lambda$4959, 5007) emission lines with asymmetric profiles, suggesting a highly asymmetric CSM.
The mid-IR continuum can be explained by $\sim${}$0.036 \ M_\odot$ of carbonaceous dust at $\sim$300~K and $\sim${}$0.043 \ M_\odot$ of silicate dust at $\sim$200~K.  
The observed dust mass has increased tenfold since the last {Spitzer} observation 4~yr ago, with evidence suggesting that new grains have condensed in the cold dense shell between the forward and reverse shocks.
This dust mass places SN\,2014C among the dustiest SNe in the mid-IR and supports the emerging observational trend that SN explosions produce enough dust to explain the observed dust mass at high redshifts. 
\end{abstract}


\section{Introduction}\label{sec:intro}
% I have to reread all of the new papers. 

% \citep{Milisavljevic2015, Margutti2017, Anderson2017, Tinyanont2019, Bietenholz2018, Sun2020, Bietenholz2021, Brethauer2022, Thomas2022, Vargas2022}

% Something about binary evolution, mass loss, stripped envelope, etc. 

Mass loss dictates the evolution of massive stars ($\gtrsim$~8~$M_\odot$) and their eventual deaths as core-collapse supernovae (CCSNe). 
For about a third of all CCSNe \citep[e.g.,][]{Smith2011rate, Shivvers2017}, the progenitor star loses its hydrogen (Type Ib) or even helium (Type Ic) envelopes and explodes as a stripped-envelope (SE) supernova \citep[SESN; for review, see, e.g.,][]{Filippenko1997, Gal-Yam2017}.
For about 10\% of CCSNe \citep{Smith2011rate, Shivvers2017}, mass loss concludes soon before, or is still ongoing, at the time of core collapse, leaving a nearby dense circumstellar medium (CSM).
% The SN shock interacts with this CSM, converting some kinetic energy into light, 
The supernova (SN) shock interacts with this CSM, heating the material and producing light, resulting in luminous interacting SNe with relatively narrow ($\sim$100--1000 $\rm km\,s^{-1}$) emission lines from hydrogen (Type IIn; \citealp{Schlegel1990}), helium (Type Ibn; \citealp{Pastorello2008}), or heavier elements (Type Icn; \citealp{Gal-Yam2021,Pellegrino2022}). 
By observing these interacting SNe, we can reconstruct the CSM structure and the mass-loss history as the SN shock sweeps out and interacts with older material. %, interacting with older ejected materials.
% The longer we monitor, the further in the past we probe.
Critically, emission from interacting SNe emerges in the infrared (IR) at late times as the interaction flux gets absorbed, thermalized, and reemitted by dust, either pre-existing or newly formed.
Therefore, late-time observations in the IR reveal the nature of the mass-loss process responsible for the diversity of the CCSN population.

%The process responsible for such a large mass removal remains debated, but several lines of observational evidence are pointing towards binary interaction \citep{Podsiadlowski1992} as the responsible mechanism for producing the majority of SESNe progenitors (\textbf{citations here}, see a review by \citealp{Smith2014}).

%is this necessary
Observations in the past decade provide evidence that binary interaction \citep{Podsiadlowski1992} is the primary mechanism for producing the majority of SESN progenitors (\citealp{Eldridge2013}; see reviews by \citealp{Smith2014, Smith2017}).
First, single-star wind-driven mass loss (even assuming high mass-loss rate classical wind prescriptions; e.g., \citealp{deJager1988}) cannot explain the relatively large SESN fraction, as only the most massive CCSN progenitors could completely lose all of their hydrogen this way. 
Empirical measurements of mass-loss rates for massive stars have also shown that the classic prescriptions overestimate the mass-loss rate by a factor of $\sim$20 \citep{Beasor2018,Beasor2020}, further limiting the prospect of stripping a star in this manner \citep{Beasor2022}. 
Second, the majority of massive stars capable of producing CCSNe live in close-in binary systems, in which mass transfer is expected \citep{Sana2012}.
Binary mass-transfer can strip lower-mass stars and could explain the high rate of SESNe \citep{Smith2011rate, Shivvers2017} and their typically low ejecta mass \citep{Drout2011, Lyman2016}. 
Recently, a population of stripped intermediate-mass helium stars thought to be produced by this process has been identified in the Magellanic Clouds \citep{Drout2023}.
Environmental studies of SESN sites point to generally intermediate age ($\sim$10 Myr), inconsistent with the very young age ($\sim$1 Myr) expected if they come from very massive stars \citep[e.g.,][]{Galbany2018, Kuncarayakti2018, Sun2020, Sun2023}, further suggesting that most SESNe come from lower-mass progenitors. 
Most recently, direct imaging with \textit{Hubble Space Telescope} (\textit{HST}) has detected putative surviving companion stars at the positions of several nearby SESNe, further supporting the binary scenario \citep[][and references therein]{Zapartas2017,Ryder2018, Fox2022}. 

To measure the mass-loss rate and CSM profile and further support the binary origin of SESNe, we need to directly observe events with CSM interaction.
Unlike other interacting SNe that start interacting immediately after the explosion, we expect CSM around SESNe to be at some distance from the progenitor;  the stripped star is expected to live for some time after the conclusion of the binary mass-transfer process.
%depending on the CSM velocity and the time delay between the conclusion of the envelope stripping phase and core collapse. 
As such, the interaction could begin months to years after the explosion, or not at all if the delay is too great and the CSM has already been dispersed by the time the progenitor dies. 
To detect these late CSM interactions, observations in the IR are key as the spectral energy distribution (SED) of the SN shifts red owing to newly formed or pre-existing dust.
% Late-time observations, especially in the IR to detect the peak of the spectral energy distribution (SED), are thus required to detect and characterize delayed interactions. 

% At these epochs, most light from the SN is in the IR due to reprocessing by dust, either pre-existing in the CSM or newly formed behind the shock and/or in the ejecta.

Massive stellar systems experiencing mass loss before undergoing a SESN, as well as interacting SN progenitors, have conditions suitable for dust production.
Dust has been observed in Galactic post-binary-interaction systems, which are thought to be SESN progenitor candidates, such as RY Scuti \citep{Gehrz1995, Smith2011RY} and NaSt1 \citep{Mauerhan2015}. 
Other extreme massive Galactic stars thought to resemble interacting SN progenitors, such as luminous blue variables (LBVs), also harbor dusty CSM \citep{Waters1998,Agliozzo2021}. 
% These dust grains reprocess interaction flux into the IR, especially at late times where 
% Likewise, 
In addition, large quantities of dust are present in interacting SNe, reprocessing the interaction flux and making them especially luminous and long lasting in the IR \citep{Fox2011, Fox2013, Tinyanont2016, Szalai2019}. 
% At late times, the IR dominates the spectral energy distribution (SED) of these transients.
Consequently, the IR dominates the SED of these transients at late times.
Thus, IR observations of interacting SNe months to years after explosion are crucial for (1) probing the presence of  detached CSM around SESNe, and (2) determining whether the CSM dust can survive the shock passage, or reform behind the shock, and be dispersed into the interstellar medium (ISM). 
%sentence on high z dust?

% A rare class of objects can help us answer both questions on mass loss process : SESNe with delayed interactions with detached CSM. 
% Owing to all-sky transient surveys, such as A, 
We have discovered at least eight SESNe, listed below, that show late-time rebrightening and spectroscopic evolution consistent with a delayed interaction with detached hydrogen-rich CSM from the lost stellar envelope.
% Observing these objects directly probes the mass loss process that creates SESNe. 
Members of this class include SNe\,2001em \citep{Chugai2006, Chandra2020}, 2004dk \citep{Mauerhan2018, Pooley2019, Balasubramanian2021}, 2014C \citep{Milisavljevic2015, Margutti2017}, and more recently SNe\,2018ijp \citep{Tartaglia2021}, 2019oys \citep{Sollerman2020}, and 2019yvr \citep{Kilpatrick2021, Ferrari2024}.
% Recently, there have been observations of other SESNe exhibiting similar interaction features, albeit weaker.
% For instance, the following SNe showed narrow hydrogen lines emerging with some delay after the explosion: S, 2019oys \citep{Sollerman2020} and 2019yvr \citep{Kilpatrick2021}.
Some objects, like SN\,2019tsf \citep{Sollerman2020, Zenati2022} and SN\,2022xxf \citep{Kuncarayakti2023}, have shown rebrightening and radio emission without hydrogen or helium lines, which have been attributed to interactions with hydrogen-free and helium-poor CSM.
In addition, there are SESNe with interactions inferred from archival IR photometry from the Wide-field Infrared Survey Explorer \citep{Myers2024}.
Among this emerging class of interacting SESNe, the most well-observed member remains SN\,2014C. 

%14C summary
% SN\,2014C presents a unique opportunity to study mass loss leading up at a SESN. 
SN\,2014C was first discovered on 2014 January 5 (UTC dates are used throughout this paper) \citep{Zheng2014} in NGC 7331.
We adopt the Cepheid distance of $14.7 \pm 0.6$ Mpc to the host galaxy \citep{Freedman2001}. 
The recent photometric analysis by \citet{Zhai2025} puts the explosion date on 2014~January~1 (MJD 56658.91).
However, we keep January 5 as our reference epoch to be consistent with \citet{Tinyanont2019}, as the difference of 4 days is insignificant at our epoch of observation. 
The SN was quickly classified as a hydrogen-poor Type Ib SN \citep{Kim2014}. %, showing clear helium features and no hydrogen.
After a few months in solar conjunction, it reemerged and was observed {on 2014 April 20, 105} days post-explosion with growing narrow- and intermediate-width hydrogen emission lines \citep{Zhai2025}, marking a transition from Type Ib to Type IIn.
The shocked CSM produced strong radio and X-ray emission \citep{Margutti2017}.
These data highlighted an interaction between the SN shock and the hydrogen-rich envelope lost from the progenitor star. 
% , the SN shock crashed into the lost hydrogen-rich envelope, emitting strongly in the X-ray and radio, and producing relatively narrow H lines like a typical SN IIn \citep{Milisavljevic2015, 
% At the time, such an interaction was not commonly observed among SNe Ib/c, with only a few sparsely observed examples like SNe\,2001em \citep{Chugai2006, Chandra2020} and 2004dk 
\citet{Margutti2017} presented archival radio observations and showed that SN\,2014C-like interaction was observed in $\sim$10\% of SESNe.
% We note that more early-time optical photometric and spectroscopic data have been recently published by \citet{Zhai2025}, suggesting that weaker interaction might have commenced as early as 20 days post-explosion.
Recent optical data published by \citet{Zhai2025} suggest that weaker interaction may have commenced as early as 20 days post-explosion.

Early time observations, particularly at X-ray wavelengths, reveal that SN\,2014C has a dense, uniform CSM shell with $\sim 1\,M_{\odot}$ at about $5\times 10^{16} \, \rm cm$ (3000 au) away from the star \citep{Margutti2017}.
More recently, \citet{Mauerhan2018}, \citet{Thomas2022}, and \citet{Brethauer2022} presented long-term observations of SN\,2014C in the optical and X-rays. 
% The optical line profiles evolve slowly, indicating a smooth CSM.
Their key finding is that the CSM must be asymmetric, with high-density regions required to explain different line widths observed in H$\alpha$ and various metal lines, and the low density inferred from X-ray observations. 
In addition, very-long-baseline interferometry (VLBI) of SN\,2014C spatially resolved the forward shock, showing minimal deceleration and an asymmetric shape \citep{Bietenholz2018, Bietenholz2021}.
These observations point to a toroidal CSM around SN\,2014C, and recent theoretical 3D simulation work by \citet{Orlando2024} supports this interpretation.  

% resolving an asymmetric forward shock, They propose that the CSM around SN\,2014C is confined in a torus.
To accurately measure the density profile of the CSM around SN\,2014C, we needed to measure its bolometric luminosity evolution, which required observations near the peak of the SED in the IR. 
SN\,2014C was monitored in the IR from explosion to 1920 days post-explosion as part of the SPitzer InfraRed Intensive Transients Survey \citep[SPIRITS;][]{Tinyanont2016, Kasliwal2017}.
% \cite{Tinyanont2016} show that SN\,2014C is uniquely luminous in the IR among the sample of CCSNe closer than 20 Mpc.
\cite{Tinyanont2019} modeled the bolometric light curve with a semianalytic model of \citet{Moriya2013} and showed that the CSM has a $\rho \propto r^{-2}$ density profile with a large mass-loss rate of $\sim 10^{-3} \ M_\odot\, \rm yr^{-1}$,  consistent with binary-induced mass loss. 
With a 10 $\mu$m detection from the ground, they also showed that silicate dust, about 30\% by mass, must be present in the CSM of SN\,2014C. 
The inferred dust mass, assuming the mixed carbonaceous and silicate composition, remained constant around $5\times 10^{-3} \ M_\odot$ throughout the observations, indicating that the dust was pre-existing. 
By the end of the {Spitzer} mission in 2020, SN\,2014C remained a luminous IR source.

Here, we present {James Webb Space Telescope (JWST)} observations of SN\,2014C 3477--3568 days post-explosion, along with ground-based optical and near-IR (NIR) spectroscopy from similar epochs.
We describe the observations and data reduction in Section~\ref{sec:obs}. The dust parameter fitting and bolometric luminosity are presented in Section~\ref{sec:analysis}. Section~\ref{sec:spec} discusses the evolution of spectral line profiles.
Our conclusions are summarized in Section~\ref{sec:conclusion}.

\begin{figure*}[!ht]
    \centering
    \includegraphics[width=\linewidth]{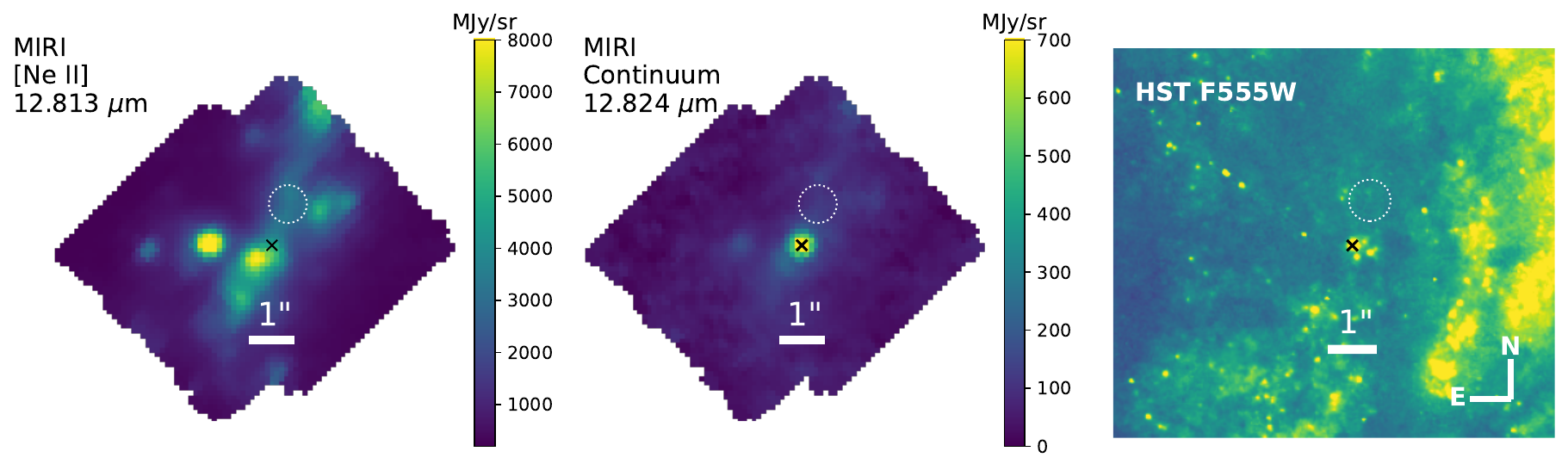} \\
    \includegraphics[width=0.8\linewidth]{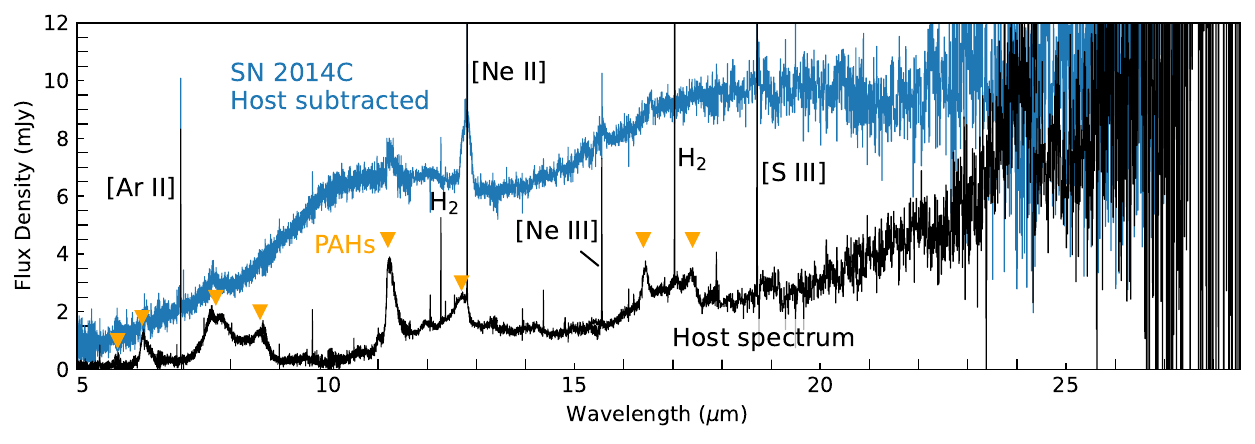}
    \caption{\textbf{Top:} images of SN\,2014C from MIRI MRS slice at the rest wavelength of [Ne~II] 12.813 $\mu$m (left), continuum at 12.824 $\mu$m (center), and {HST}/WFC3 F555W (right; PID 16691; PI Foley).
    The SN is marked with a cross. 
    North is up, and east is to the left in all images.
    The {HST} image was used to align the MIRI parallel image, and applied to the MIRI MRS data. 
    The [Ne~II] 12.813 $\mu$m image is representative of the complex host environment around SN\,2014C. 
    We note that a broad [Ne~II] line associated with the SN is detected in addition to the host emission. 
    The contaminating flux is mostly in narrow lines, and does not affect our measurements. 
    To estimate the continuum host emission, we use the circular aperture shown as a dashed white circle, which selects an area with emission similar to that at the SN location, and is far enough as to not self-subtract the SN at longer wavelengths (the SN flux in this background aperture is $<$1\% in MRS Channel 4).  
    \textbf{Bottom:} MRS spectrum of SN\,2014C with the host emission (estimated from the aperture shown in the top figures) subtracted. 
    The host spectrum used for subtraction is shown in black, in the same units as the SN spectrum.  
    {Strong narrow host lines are labeled, while PAHs bands at 5.7, 6.2, 7.7, 8.6, 11.2, 12.7, 16.4, and 17.4 $\mu$m are marked with orange triangles.}
    }
    \label{fig:14C_image}
\end{figure*}

\section{Observations and Data Reduction} \label{sec:obs}
\subsection{JWST NIRSpec and MIRI MRS}
SN\,2014C was observed with {JWST} as part of program GO-2348 (PI Tinyanont) on 2023 July 25 (3477 days post-explosion, rest frame) using the Mid-InfraRed Instrument (MIRI) in the Medium Resolution Spectrometer (MRS; \citealp{Wells2015, Argyriou2023}),  and on 2023 October 24 (3568 days post-explosion) using the Near-InfraRed Spectrograph (NIRSpec) in the integral-field unit (IFU) mode \citep{Jakobsen2022, Boker2023}. 

% The MIRI MRS observations covered 4.9 to 27.9 $\mu$m continuously, using three gratings. 
We obtained MIRI MRS observations in all subbands to ensure the full wavelength coverage of 4.9 to 27.9 $\mu$m.
We used a four-point dither pattern optimized for point sources to observe the SN. 
All exposures used the FASTR1 readout pattern with 30 groups per integration, 1 integration per exposure, and 1 exposure at each dithering point (4 total), yielding the total of 333~s exposure time for each subband. 
Dedicated background observations were also performed with the same exposure time but without dithering.
We also obtained parallel MIRI imaging in the F560W, F1000W, and F1130W filters, covering the field adjacent to the SN. 

The NIRSpec IFU observations were conducted with two filter/grating configurations, F170LP/G235H and F290LP/G395H, to cover 1.66--5.27 $\mu$m.
Both observations used the four-point-nod dither and the NRSIRS2RAPID readout pattern.
The F170LP/G235H and F290LP/G395H observation employed five and three groups per integration, respectively. 
Both observations used one integration, and one exposure at each dither point (four total), yielding 350 and 233~s total exposure times for the F170LP/G235H and F290LP/G395H setups, respectively.
Owing to the expected brightness of the source, no ``leakcal'' observations to remove contamination from stuck-open shutters in the microshutter assembly were requested. 

From the automatically reduced spectral cube provided on the Barbara A. Mikulski Archive for Space Telescopes (MAST), we note that the observations are offset from previous {HST} observations of the same field.\footnote{These data can be retrieved at \url{http://dx.doi.org/10.17909/adcj-n781}} 
To improve the world coordinate system (WCS) solution of the final data cube, we measure the coordinate offsets in our observations by the following procedure. 
First, we download an {HST} image of the field that covers the SN and has a significant overlap with the MIRI parallel images. 
We used the F555W image obtained with the Wide-Field Camera 3 (WFC3) on 2022 July 31 (PID 16691; PI Foley; this image along with the aligned MRS cube is shown in Figure~\ref{fig:14C_image}, top right). 
We align it to the {Gaia} DR2 catalog using the {JWST} {HST} Alignment Tool (JHAT; \citealp{Rest2023_jhat})\footnote{\url{https://jhat.readthedocs.io/}} and produce a secondary catalog of sources from the deeper {HST} image. 
We then align the MIRI parallel image in the F560W band to this secondary catalog. 
We use the bluest MIRI image to ensure that there are enough common point sources between it and the optical image. 
% After this process, the 
Finally, we compute the R.A. and decl. offsets from this process, and apply the same offset to the coordinates in the MRS files.

We use the {JWST} data-reduction pipeline v1.15.1\footnote{\url{https://jwst-pipeline.readthedocs.io/}} to process the raw data with the Calibration Reference Data System (CRDS) v11.17.20.\footnote{\url{https://jwst-crds.stsci.edu/}} 
For both NIRSpec and MIRI MRS, we download uncalibrated files from MAST and run the Stage 1 pipeline to fit the observed ramp and produce rate files. 
The reference coordinates for the rate files are updated by adding the R.A. and decl. offsets found using JHAT to the \texttt{RA\_REF} and \texttt{DEC\_REF} FITS header keywords of MRS observations.
We run the \texttt{spec2} pipeline stage to generate calibrated files. 
At this stage, we remove the background from the telescope and the zodiacal light.
We extract 1D background spectra from our dedicated background pointing using the \texttt{Extract1D} function with the source type set to \texttt{EXTENDED}.
This way, the entire field of view is used, reducing the noise in the measurement. 
We then rerun the \texttt{spec2} pipeline using this extracted spectrum for master background subtraction. 

%data cubes from each MRS and NIRSpec IFU observation. 
After this step, we run the \texttt{spec3} pipeline stage to drizzle and combine the dithered, dedicated-background-subtracted observations, keeping different channels and bands separate.  
This results in 12 spectral cubes from MRS, from Channels 1 to 4, with the short, long, and medium bands per each channel. 
% For MRS, we use the 1-dimensional extracted background spectra from the dedicated background observation for background subtraction in this stage. 
% We then extract 1-dimensional background spectra from the dedicated background observations and use them for our master background subtraction. 

% Finally, the pipeline extracts 1D spectra for the source by performing aperture photometry in each slice. 
% We set the aperture size to be equal to the FWHM of the PSF at that wavelength (by setting the parameter \texttt{ifu\_rscale} $=1.0$).

\subsection{Host Background Subtraction}
While the thermal background from the sky and the telescope is removed using the dedicated background observations, the cubes still have a significant, spatially varying host-galaxy background underneath the SN. 
% To remove this contribution, we use a post-processing script AstroBkgInterp 
% \textbf{get text from Bryony and Mike.}
% Estimating this background component is complicated. 
The pipeline provides a basic tool to measure and subtract background in an annulus around the source during spectral extraction, and the annulus size scales with the wavelength. 
% This tool is insufficient because the annulus radius and size scale with 
However, with our spatially varying background, it is crucial to sample the background from the same region at all wavelengths. 
To do this, we place a circular aperture with a radius of 0\farcs52 at $\alpha = 22^{\rm h} 37^{\rm m}05.57^{\rm s}$, $\delta = +34^\circ 24' 32\farcs517$, shown in Figure~\ref{fig:14C_image} (top). 
This location is along the prominent elongated structure running northwest to southeast through the SN, and it is far enough away from the SN to include $< 1\%$ of the SN light at the reddest wavelength according to the point-spread-function (PSF) model. 
It also avoids all the star-forming knots observed in strong lines (e.g., Figure \ref{fig:14C_image}, top left; other lines show star-forming knots at similar locations).  
At each wavelength slice in the spectral cubes, we subtract the surface brightness measured in this aperture from the whole slice.

Finally, we perform spectral extraction of the SN from the host-subtracted cubes using the \texttt{Extract1d} step in the {JWST} pipeline.
Figure~\ref{fig:14C_image} (top) shows the final spectral cube from MRS compared with the {HST} image. 
The brightest source in the MRS cube is well aligned with the location of the SN from the {HST}/WFC3 F555W image from 2022 (PID 16691; PI Foley). 
We extract the 1D spectrum at this location from the host-subtracted cubes for further analysis.
The {JWST} pipeline extracts 1D spectra for a point source by performing aperture photometry in each slice. 
We set the aperture size to be equal to the full width at half-maximum (FWHM) intensity of the PSF at that wavelength (by setting the parameter \texttt{ifu\_rscale} $=1.0$).
An aperture correction is then applied. 
We note that fluxes in the overlapping regions between different channels and bands agree with each other, and we do not apply further scaling.
Figure~\ref{fig:14C_image} (bottom) shows the final SN spectrum along with the host background spectrum used for the subtraction. 
The strong, broad emission features from polycyclic aromatic hydrocarbons (PAHs) at (for example) 7.7, 11.3, and 16.4 $\mu$m, which are from the host galaxy and unlikely from the SN, have been mostly removed, demonstrating that the selected background aperture is appropriate. 
Figure~\ref{fig:spec_comp} (left) compares this spectrum to the 2019 observations of SN\,2014C from \citet{Tinyanont2019} along with other objects in the literature.

% Typically, one would measure the surface brightness from an annulus to the source that has insignificant amount of light from the tail of the PSF. 
% Ideally, this background region should be the same for all wavelength slice, given the spatial variation in the host background emission.
% However, for MRS, PSF at the red end is a factor of $\sim$5 larger than in the blue end. 
% % Second, the host background has strong spatial variation. 
% Thus, we cannot sample the sa

\subsection{Keck LRIS and NIRES Optical to NIR Spectroscopy}
\begin{figure*}[ht!]
    \centering
    \includegraphics[width=1\linewidth]{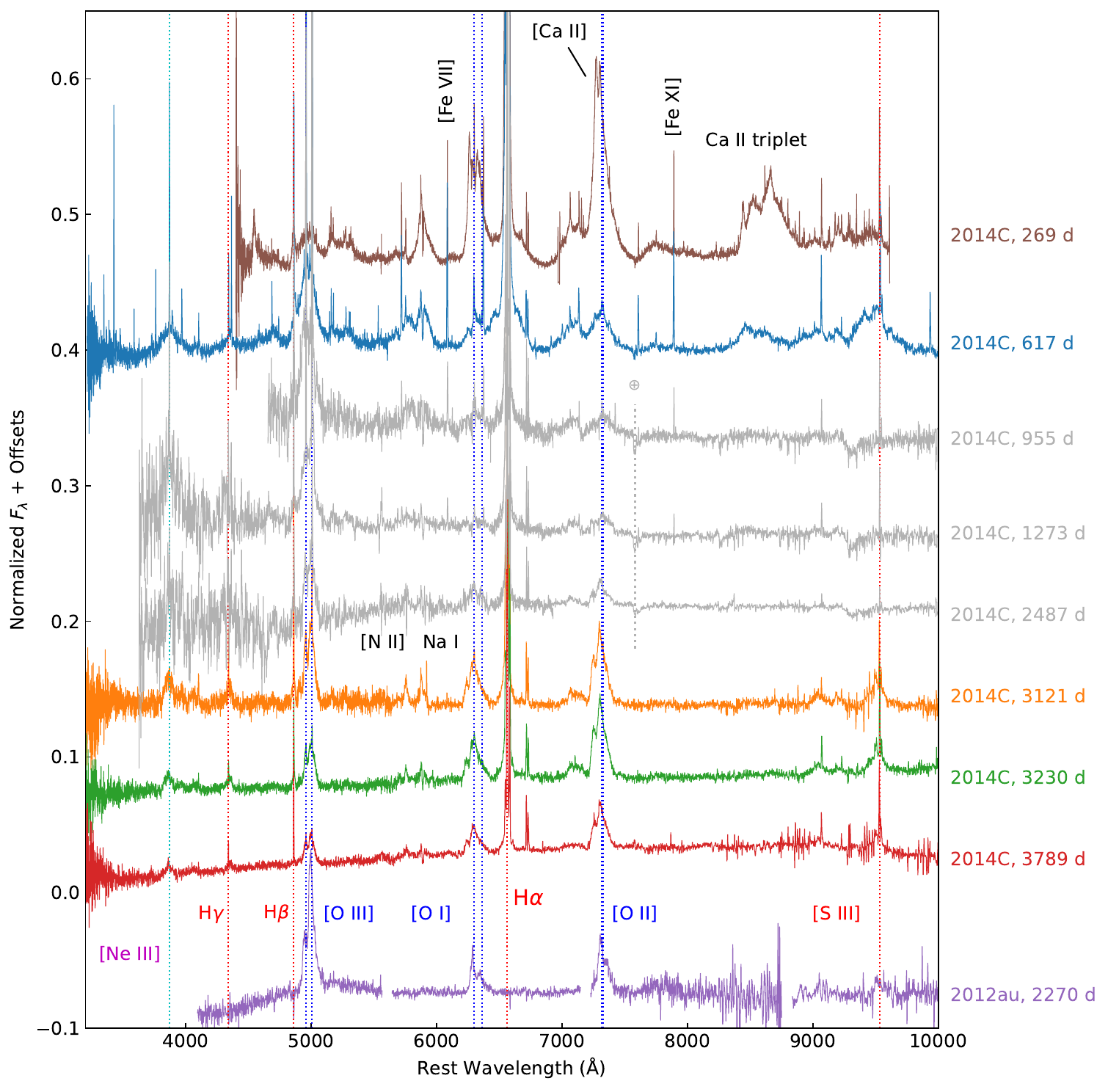}
    \caption{New optical spectra of SN\,2014C from 3131, 3230, and 3789 days compared with earlier spectra from \citet{Mauerhan2018} at 269 and 619 days, {and from \citet{Thomas2022} at 955, 1273, and 2487 days (note that we merge their red and blue spectra taken at similar time to cover a wider wavelength coverage and report the average epoch)}. 
    The spectrum of SN\,2012au at 2270 days from \citet{Milisavljevic2018} is provided for comparison.
    }
    \label{fig:new_spec}
\end{figure*}

SN\,2014C was observed with the Low-Resolution Imaging Spectrometer (LRIS; \citealp{oke1995}) on the Keck~I 10\,m telescope on Maunakea, Hawaii, on 2022 August 2 (3121 d post-explosion), 2022 November 20 (3230 d), and 2024 June 3 (3789 d).
% The data were reduced using a custom open-source \texttt{pyraf}-based pipeline.\footnote{The pipeline is available at \url{https://github.com/msiebert1/UCSC_spectral_pipeline}.}
% We used the pipeline to perform bias subtraction, flat fielding using dome flats, and wavelength calibration using observations of arc lamps.
The data were reduced using \texttt{LPipe} \citep{Perley2019}.
Flux calibration was performed using a spectrophotometric standard star observed on the same night. 
These new optical spectra are shown in Figure~\ref{fig:new_spec}.

SN\,2014C was observed with the Near-InfraRed Echellette Spectrometer \citep[NIRES;][]{wilson2004} on the Keck~II 10\,m telescope on 2024 June 21 (3807 d post-explosion) as part of the Keck Infrared Transient Survey (KITS; \citealp{Tinyanont2024}).
The observations were performed with two sets of the ABBA dithering pattern to sample the sky background, with a total exposure time of 2400\,s.
The A0\,V star HIP111538 was observed immediately before the SN to provide flux and telluric calibration. 
We reduced the data using \texttt{pypeit} \citep{prochaska2020, pypeit2020} following the procedure outlined by \citet{Tinyanont2024},  automatically performing flat-fielding, background subtraction, and source detection and extraction. 
The science spectra were then flux calibrated, coadded, and corrected for telluric absorption, using the aforementioned A0\,V star observation.
Only the \ion{He}{1} 1.083\,$\mu$m was detected, and is shown in Figure~\ref{fig:forbiddenO}.

\subsection{Subaru COMICS Mid-IR Photometry}
\begin{figure*}
    \centering
    \includegraphics[width=0.49\linewidth]{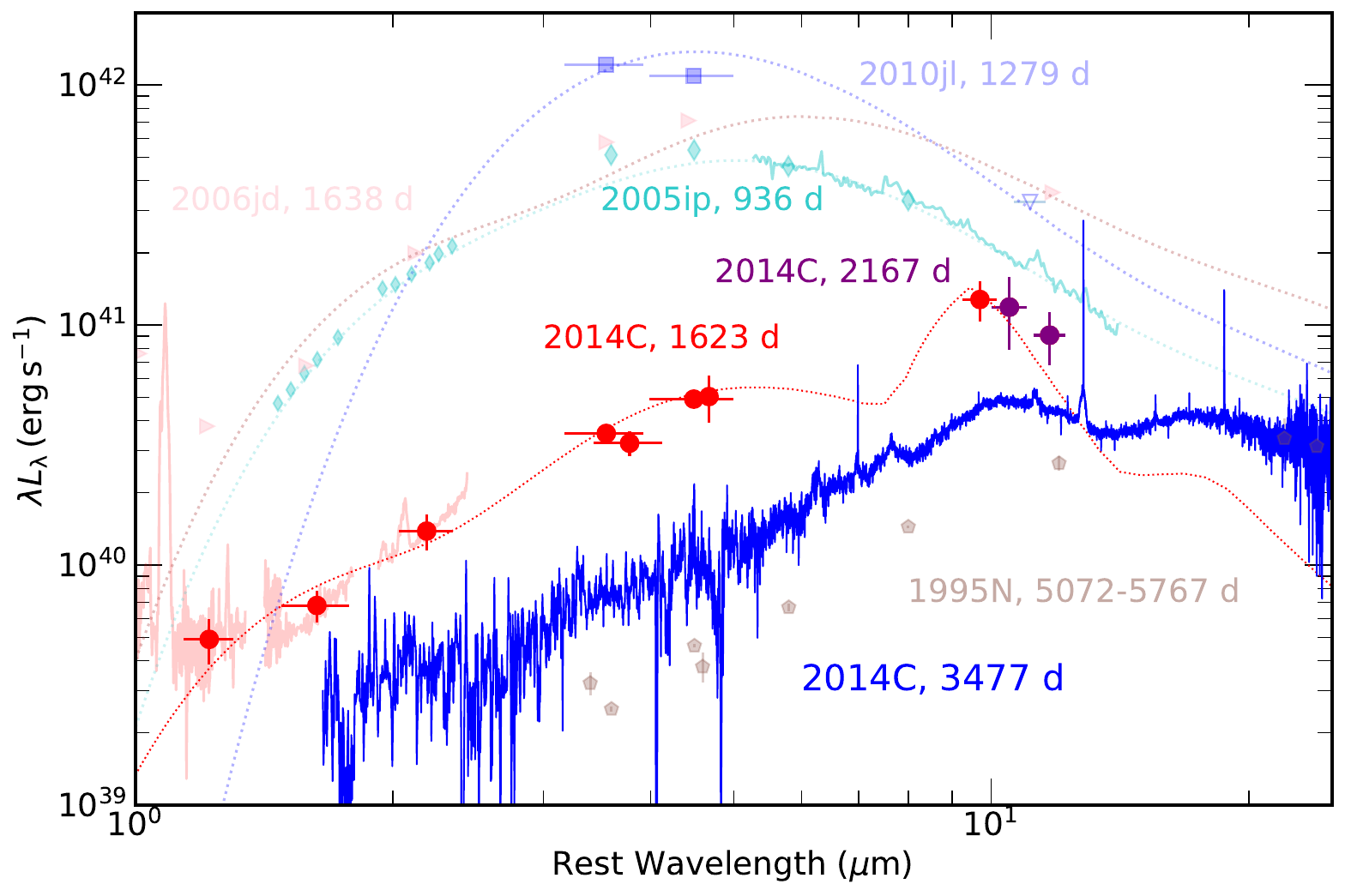} \hfill
    \includegraphics[width=0.49\linewidth]{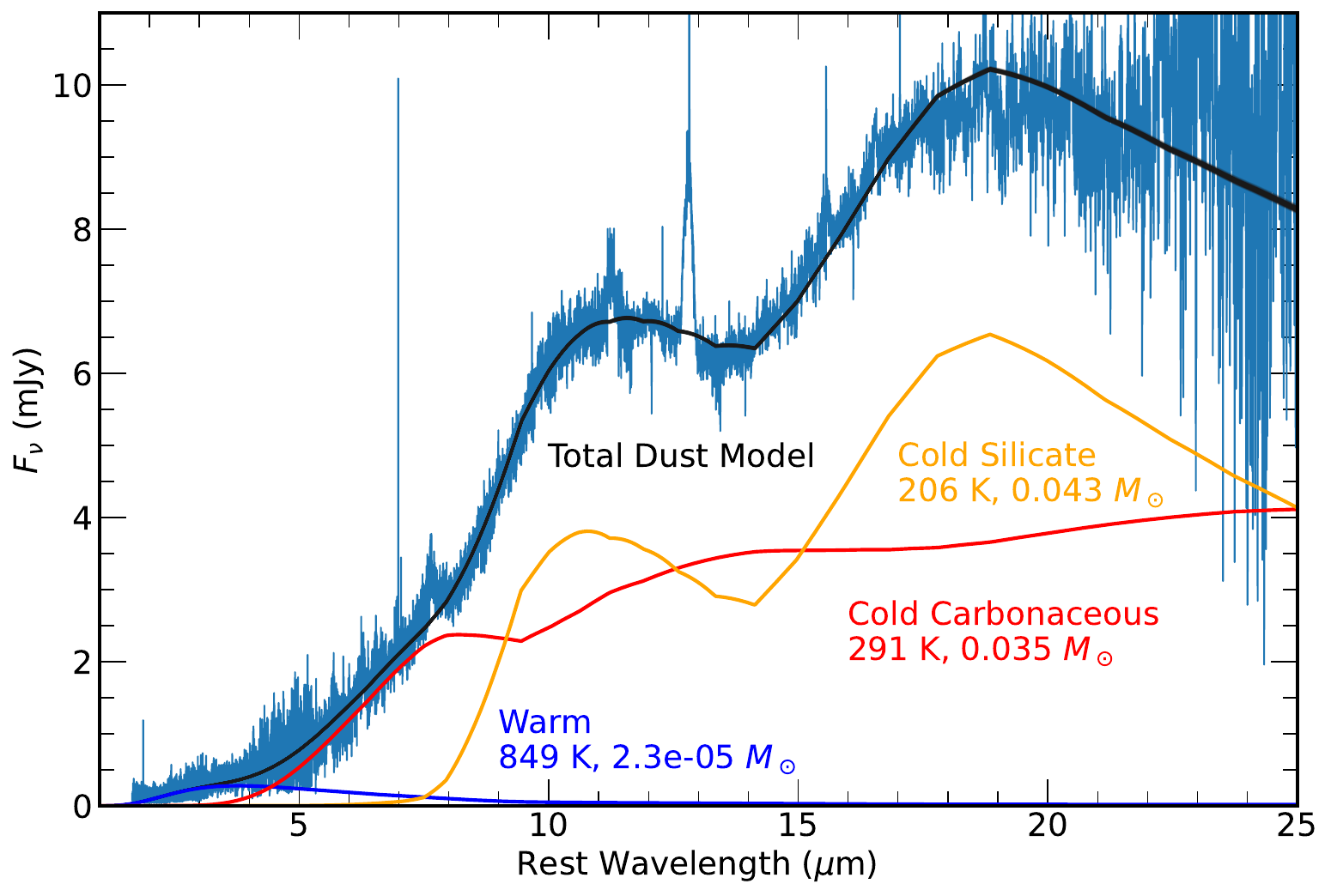}
    \caption{\textbf{Left:} NIR to mid-IR spectrum of SN\,2014C from {JWST} NIRSpec IFU and MIRI MRS, from 3568 and 3477 days post-explosion, respectively. 
    The IR spectra and photometry of SN\,2014C at 1623 days published by \citet{Tinyanont2019}, along with all mid-IR data of SNe~IIn from before 2019 (also shown in Figure~6 of \citealp{Tinyanont2019}), are displayed for comparison. 
    The ground-based mid-IR photometry from Subaru/COMICS at 2167 days post-explosion is also shown. 
    The best dust model is shown for every SN, except SN\,1995N, for which the spectrum resembles a power law.
    {For SN\,2014C, we note that the photometry at 2167 d deviates from the model fitted to the 1623-d SED. This is consistent with the SN cooling down in the 1.5 years between the two observations.}
    \textbf{Right:} mid-IR spectrum of SN\,2014C with the best-fit dust model retrieved from MCMC fitting. The model requires three dust components, whose best-fit parameters are summarized in Table~\ref{tab:dust_params}. 
    }
    \label{fig:spec_comp}
\end{figure*}

SN\,2014C was observed by the Cooled Mid-infrared Camera and Spectrometer (COMICS; \citealp{kataza2000}) on the Subaru Telescope in the N10.5 ($\lambda_\mathrm{c}=10.5$ $\mu$m, $\Delta\lambda=1.0$ $\mu$m) and N11.7 ($\lambda_\mathrm{c}=11.7$ $\mu$m, $\Delta\lambda=1.0$ $\mu$m) filters on 2019 December 12, before the instrument was decommissioned. 
%N9.7 filter ($\lambda = 9.7$ $\mu$m, $\Delta\lambda=0.9$ $\mu$m) on 2018 June 28. 
% {Ryan Lau, can you check this?}
Individual 200\,s exposures were taken in chopping-only mode with a chop amplitude of 10\arcsec; the total integration times on SN\,2014C were 94 and 37 minutes for the N10.5 and N11.7 filters, respectively. 
11~Lac (HR~8632) was used as the photometric standard star from the list of mid-IR standards given by \cite{cohen1999}. 
The measured fluxes were $16.1\pm5.4$ mJy and $13.7\pm 3.4$ mJy in the N10.5 and N11.7 filters, respectively, and are shown in Figure~\ref{fig:spec_comp} (left).
%, confirming the presence of silicate dust at that epoch. 

% \begin{figure*}
%     \centering
%     % \includegraphics[width=0.8\linewidth]{emcee_fit_3-comp_CCCSi_opt_depth.pdf}
    
%     \caption{ }
%     \label{fig:dust_fit}
% \end{figure*}

% \section{Analysis and discussion}\label{sec:analysis}
\section{Dust Emission in the Mid-IR}\label{sec:analysis}
\subsection{Spectrum Fitting and Dust Parameter Estimation}\label{sec:dust}

We fit the IR spectrum of SN\,2014C with a combination of dust models to determine the dust temperatures, masses, and compositions. 
The set of equations presented by \cite{Shahbandeh2023} is employed. 
The observed dust flux is given by
\begin{equation}
    F_{\rm dust}(\lambda) = \frac{B(\lambda, T_{\rm dust}) \kappa(\lambda) M_{\rm dust} P_{\rm esc}(\tau) }{d^2},
\end{equation}
where $B$ is the Planck function, $d$ is the distance to the SN, and $\kappa$ is the dust opacity from \citet{Draine1984} and \citet{Laor1993} assuming a grain size of $a = 0.1 \ \mu$m.
We note that, in the IR, $\lambda \gg a$, and the spectral shape is insensitive to the grain size \citep[e.g.,][]{Fox2010, Sarangi2022}.  
Large grains with $a > 1 \ \mu$m are not expected in SNe at this phase, and they cannot reproduce the strong silicate feature observed.
% and $P_{\rm esc}(\tau)$ is the escape probability at a given optical depth $\tau$. 
Per \citet{Cox1969}, the escape probability $P_{\rm esc}(\tau)$, assuming spherical symmetry, is given by 
\begin{equation}\label{eq:P}
P_{\rm esc}(\tau) = \frac{3}{4\tau} \left[1- \frac{1}{2\tau^2} +
                                        \left(\frac{1}{\tau}+\frac{1}{2\tau^2} \right) e^{-2\tau}
                                    \right].
\end{equation}
The optical depth, assuming that the dust is evenly distributed inside the expanding SN, is 
\begin{equation}\label{eq:tau}
    \tau(\lambda, t) = \rho(t) R(t) \kappa(\lambda) = \frac{3}{4} \frac{M_{\rm dust}(t) }{\pi R(t)^2} \kappa(\lambda).
\end{equation}
% We adopt the shock velocity of 10{,}000 $\rm km\,s^{-1}$, from early-time optical lines \citep{Milisavljevic2015}, to compute the radius $R = 3\times 10^{17} \ \rm cm$ at the epoch of MRS observation. 
We leave the radius $R$ as a free parameter because the shock velocity at this phase is unclear. 
We note that a generic shock velocity of 10{,}000 $\rm km\,s^{-1}$ used by \citet{Tinyanont2019}, consistent with early time spectroscopy \citep{Milisavljevic2015, Zhai2025}, would result in $R = 3\times 10^{17} \ \rm cm$ at the epoch of the MRS observation.
% Here, we note that while the spherical geometry may not be appropriate for the interacting SN\,2014C, we will later show that the optical depth is low regardless of the assumed geometry.

Here, we note that the spherical geometry assumption does not significantly affect our results. 
Because both $M_{\rm dust}$ and $R$ are free parameters, $\tau$ is effectively a free parameter as well, and the geometry is only used to compute the escape probability.
% \citep{Dwek2024} presents 
% For SN\,2014C, the observed dust is likely either in a shell ahead of the shock in case of pre-existing dust or behind the shock in case of newly-formed dust in the cold dense shell (CDS) between the forward and reverse shock. 
% Further, the CSM in SN\,2014C is likely confined in an equatorial disk \citep[e.g.][]{Bietenholz2021, Thomas2022}.
% In these scenarios, the dust is confined to a smaller volume and will have a larger optical depth than if it were evenly distributed. 
% This will lead to an even higher dust mass, given the observed flux.
As we will show, the optical depth in the IR is low at this epoch, and the assumed geometry here does not affect our results. 

% We will revisit the dust geometry in Section \textbf{XX}. 

We first determine how many dust components are required to explain our data by performing a nonlinear least-squares minimization fit using the \texttt{curve\_fit} routine from the \texttt{scipy.optimize} package. 
Each component has a unique temperature, mass, and composition (either carbonaceous or silicate, which have different opacity, but with a fixed grain size). 
The optical depth used in the fit is the total optical depth from all components considered. 
More components are added iteratively until a satisfactory fit is achieved.
We find that three dust components are required; their properties, from the subsequent Markov Chain Monte Carlo (MCMC) fitting, are summarized in Table~\ref{tab:dust_params}.
The warm component with $T\approx 800$ K is needed to explain the NIR flux, while the majority of dust mass is found in the cold components.
% The broad spectral features observed at around 11 and 18 $\mu$m necessitate some silicate dust.
The broad spectral features observed at around 11 and 18 $\mu$m are indicative of a cold silicate dust component.
Another cold carbonaceous component is needed to fit the overall flux in the mid-IR.

%This is with fixed r, v = 10000 km/s.
% \begin{table}[]
%     \centering
%      \caption{Best-fit dust parameters for SN\,2014C}
%     \begin{tabular}{cccc} \toprule
%     Component & $T$ (K) & $M$ ($M_\odot$) & Composition \\ \hline
%      % Hot     & $2229\pm 9$ & $(1.37 \pm 0.02)\times10^{-6}$ & C \\
%      Warm    & $822\pm 3 $ & $(2.74 \pm 0.07)\times10^{-5}$ & C \\
%      Cold C  & $295\pm 1 $ & $0.0316 \pm 0.0003$ & C \\
%      Cold Si & $216\pm 1 $ & $0.0402 \pm 0.0002$ & Si \\ \hline 
%      Total & 251$^\dagger$ & $0.0718 \pm 0.0004 $ & \\ \hline 
%     \end{tabular}
%     \label{tab:dust_params}
%     \vspace{6pt}
%     {\raggedright $^\dagger$ Mass-averaged temperature \par}

% \end{table}

%This is with free R. 
\begin{table}[]
    \centering
     \caption{Best-fit dust parameters for SN\,2014C}
    \begin{tabular}{cccc} \toprule
    Component & $T$ (K) & $M$ ($M_\odot$) & Composition \\ \hline
     % Hot     & $2229\pm 9$ & $(1.37 \pm 0.02)\times10^{-6}$ & C \\
     Warm    & $850\pm 4 $ & $(2.30 \pm 0.05)\times10^{-5}$ & C \\
     Cold C  & $291\pm 1 $ & $0.0355 \pm 0.0005$ & C \\
     Cold Si & $207\pm 1 $ & $0.0425 \pm 0.0003$ & Si \\ \hline 
     Total & 245$^\dagger$ & $0.0780 \pm 0.0006 $ & \\ \hline 
    \end{tabular}
    \label{tab:dust_params}
    \vspace{6pt}
    {\raggedright $^\dagger$ Mass-averaged temperature. \\
    Uncertainties provided are statistical errors only. The mass uncertainty is dominated by the distance uncertainty to the SN. \par
    }

\end{table}

To sample the posterior distribution of our fitted parameters, we perform an MCMC fit using \texttt{emcee} \citep{emcee}.  
We assume a flat prior distribution for all parameters within a physical range, and use the results from our preliminary least-squares minimization fit as initial values. 
Further details of the MCMC run can be found in the Appendix, and the results including 1$\sigma$ uncertainties are summarized in Table~\ref{tab:dust_params}. 
The mass uncertainty provided here only includes statistical uncertainty.
A more representative uncertainty for dust mass is twice the distance uncertainty to the SN: about 8\%. 
The best-fit radius (assuming spherical geometry) is $R = (3.82\pm0.06) \times 10^{17} \rm \ cm$.
% We reiterate that we leave $R$ as a free parameter to effec
The best-fit model is shown over the data in Figure~\ref{fig:spec_comp} (right).

% From the best-fit masses and radius, we compute the optical depth for all dust components and find that the optical depth of the cold silicate and carbonaceous components is $\tau \approx 0.15$ in the IR continuum.
{The average best-fit optical depth of the cold silicate and carbonaceous dust components is $\tau \approx 0.15$ in the IR continuum.}
The symmetric posterior distribution (Figure~\ref{fig:corner}) of the dust masses also indicates that the dust is optically thin; otherwise, it would plateau toward high mass as the majority of the dust cloud is not visible. 
Here, we reiterate, that by leaving $R$ as a free parameter, $\tau$ is effectively a free parameter as well, so the result is only dependent on the assumed geometry through the $P_{\rm esc}$ calculation. 
% Here, we note that if the dust distribution is concentrated in a shell or torus rather than in a sphere (which we assume), equations \ref{eq:P} and \ref{eq:tau} would be different. 
% The optical depth would be slightly higher, leading to more dust mass inferred from the observation. 
% We will also discuss in the next section that the profile of forbidden lines precludes a geometry where the dust is concentrated in a very dense shell. 

The observed mid-IR spectrum of SN\,2014C indicates that SN\,2014C now harbors a total of $0.0780 \ M_\odot$ of cold dust with a temperature of $\sim 245$\,K. 
A mixture of carbonaceous (46\% by mass) and silicate (54\%) dust is required to fit the observed spectral features. 
This is roughly consistent with the silicate dust mass fraction found previously by \citet{Tinyanont2019}.
From 2019 until 2023, the minimum dust mass required to fit the SED of SN\,2014C has increased by more than an order of magnitude (Figure~\ref{fig:LTM_evo}b), while the temperature has dropped from around 500~K to 250~K (Figure~\ref{fig:LTM_evo}a).
Such an evolution points to new dust formation, which we discuss in more details in Section~\ref{sec:dust_origin}. 
% Next, we discuss the dust location and its likely origin as newly-formed dust in the cold dense shell. 

\subsection{Bolometric Luminosity}\label{sec:bolo}
The total luminosity of the SN at this epoch is key to determining its powering mechanism, almost 10\,yr post-explosion. 
The SED of SN\,2014C at the {JWST} epoch peaks in the mid-IR, around 20 $\mu$m. 
Therefore, we measure the bolometric luminosity by integrating the high signal-to-noise ratio (S/N) part of the observed spectrum between 2.5 to 22 $\mu$m, and adding the missing flux from the best-fit dust model discussed in the previous section.
We determine the bolometric luminosity at 3477 days post-explosion to be $(6.5 \pm 0.2) \times 10^{40} \rm \ erg\,s^{-1}$ (86\% of which is observed with {JWST} and 14\% is from extrapolation).

The bolometric luminosity from the {JWST} observations is shown in Figure~\ref{fig:LTM_evo} (c), along with the bolometric light curve inferred from {Spitzer} observations \citep{Tinyanont2019}. 
The powering mechanism at this epoch is the CSM interactions that get reprocessed into the IR by dust. 
The semianalytic light curve of an SN interacting with a CSM with $\rho = D r^{-2.01\pm0.01}$, where $D = 10^{14.9 \pm 0.2} \rm \ g\,cm^{-1}$, presented by \citet{Tinyanont2019} (fit to only the {Spitzer} data), is plotted. 
The observed bolometric luminosity agrees well with the model prediction, indicating that the SN is still interacting with the same CSM component responsible for the light curve up to 1920 days post-explosion. 
The lack of extra luminosity, which would have been evidence for an additional denser CSM, rules out the possibility that the increased observed dust mass is more pre-existing dust farther out from the SN. 
% This is our best evidence that the observed dust mass in SN\,2014C at this epoch is large

% For consistency with \citet{Tinyanont2019}, we assume the shock velocity of $10{,}000 \ \rm km \, s^{-1}$, the shock radius is at $3\times 10^{17} \rm \ cm$. 
Furthermore, we can compute the mass in the extended CSM that has interacted with the SN so far, using the CSM density profile from \citet{Tinyanont2019}.
Because of the wind-like $\rho \propto r^{-2}$ profile, the swept-up CSM mass is $M_{\rm CSM, wind} = 4 \pi D r = 1.5 \substack{+0.9 \\ -0.6} \ M_\odot$, where we compute $r$ by assuming the shock velocity of $10{,}000 \ \rm km \, s^{-1}$ (to be consistent with the same calculation by \citealp{Tinyanont2019}).
If we use the best-fit radius ($3.8\times 10^{17} \rm \ cm$) from the dust-fitting process, we get $M_{\rm CSM, wind} = 1.9 \substack{+1.1 \\ -0.7} \ M_\odot$.
The takeaway is that $\sim 2 \ M_\odot$ of the CSM has been processed by the shock. 

% The amount of shocked CSM gas now in the CDS is about 200 times more than the observed dust mass, which is consistent with our proposed scenario that the dust is newly formed i

\subsection{Dust Location and Origin}\label{sec:dust_origin}
While the blackbody radius can constrain the location of the dust, we show here that it may not be sufficient to discriminate between different dust origins.
The blackbody radius is usually computed by assuming that the observed dust luminosity and temperature come from blackbody radiation. 
Here, $r_{\rm bb} = \sqrt{L/4\pi \sigma_{\rm SB} T^4}$, where $\sigma_{\rm SB}$ is the Stefan--Boltzmann constant. 
If the dust cloud is optically thin in the IR, then $r_{\rm bb}$ is the lower limit of the dust location; if the IR optical depth is much smaller than 1, $r_{\rm bb}$ is not a constraining lower limit.
Moreover, this analysis assumes that the dust distribution is spherically symmetric, which may not be the case in most interacting SNe, including this SN.

\begin{figure*}
    \centering
    \includegraphics[width=\linewidth]{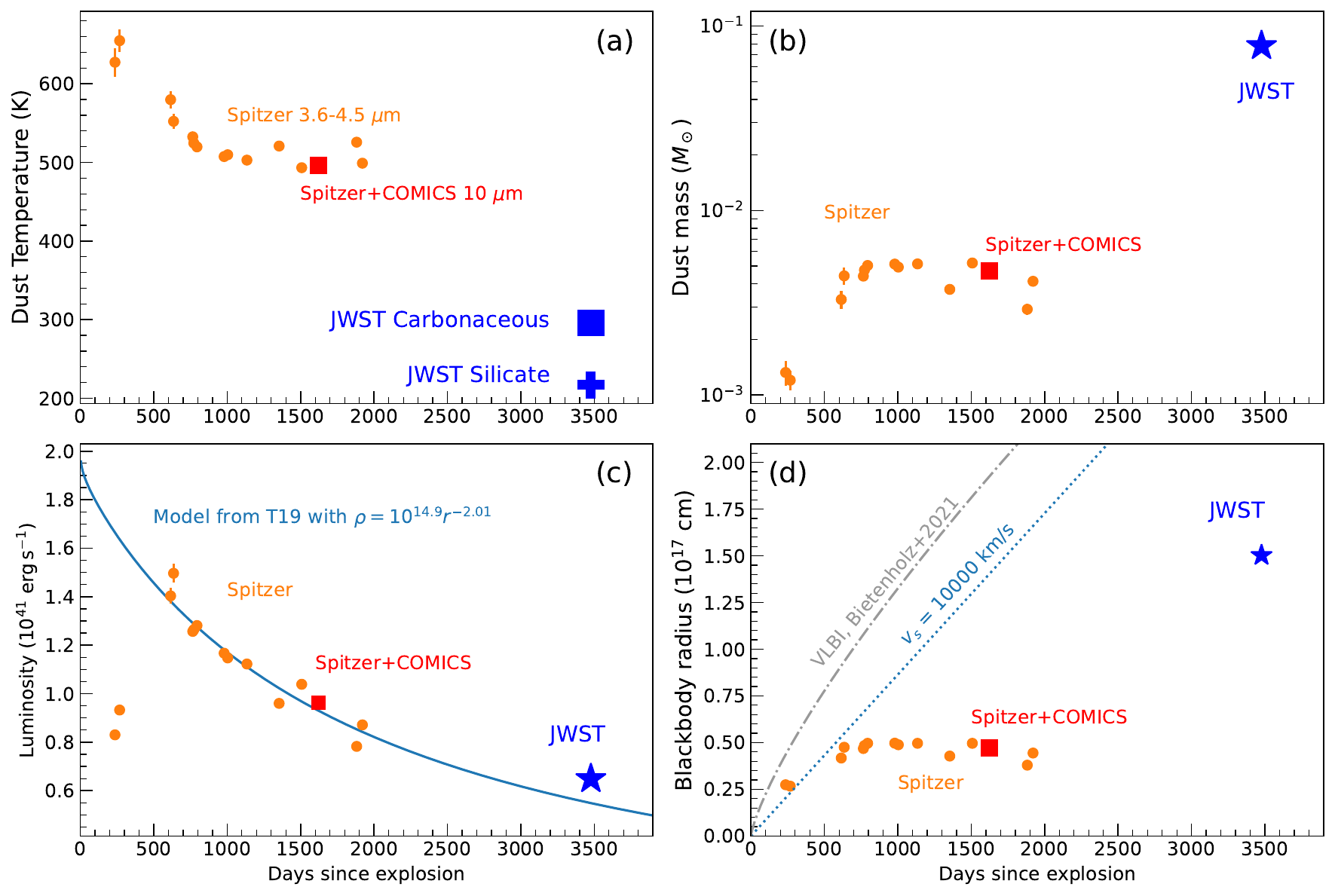}
    \caption{
    \textbf{(a)} Dust-temperature evolution of SN\,2014C. 
    Orange points are from {Spitzer} 3.6 to 4.5 $\mu$m observations, and the red point is from {Spitzer} plus a ground-based 9.7 $\mu$m image from \citet{Tinyanont2019}.
    These measurements are photometric, and the dust models  have a mixture of carbonaceous and silicate dust at a single temperature. The mass ratio is fixed by the observations including the 10 $\mu$m point. 
    The blue square and plus sign are the carbonaceous and the silicate dust components observed by {JWST}.
    \textbf{(b)} Dust mass evolution, showing an order of magnitude jump from $\sim 5 \times 10^{-3} \ M_\odot$ during the {Spitzer} era to 0.078 $M_\odot$ observed by {JWST} (total of both carbonaceous and silicate dust). We note that this is the minimum mass required to fit the observed SED. 
    \textbf{(c)} Bolometric luminosity. The semianalytic model fit to the {Spitzer} data is plotted as a cyan line (from \citealp{Tinyanont2019}). 
    The new observation shows that SN\,2014C is behaving as expected by the semianalytic model, indicating that the interaction with the same wind-like CSM component is still ongoing.    
    \textbf{(d)} Blackbody radius of the IR emission. Two estimates of the shock location are provided: one derived from the VLBI measurement \citep{Bietenholz2021}, and one with a constant velocity of 10,000 $\rm km\,s^{-1}$. 
 {We note that error bars are plotted for all points, but are smaller than the marker in some cases.}
    }
    \label{fig:LTM_evo}
\end{figure*}

We compute the blackbody radius at the {JWST} epoch, using the total luminosity and the mass-weighted averaged temperature of the dust since we have two dominating components. 
Using this temperature only affects the final radius by less than a percent. 
The computed blackbody radius is $r_{\rm BB} = (1.50 \pm 0.02) \times 10^{17} \ \rm cm$.  
This is about half of the current shock radius ($3\times 10^{17} \rm \ cm$), assuming the shock velocity of $v_s = 10{,}000 \rm \ km\,s^{-1}$. 
Figure~\ref{fig:LTM_evo} (d) shows the evolution of the blackbody radius inferred from IR observations of SN\,2014C. 

% Since we have shown in the last section that the dust emission in SN\,2014C at the JWST epoch is optically thin, the blackbody radius serves as a lower limit of the size of the dusty sphere.
However, the IR optical depth of the dust emission in SN\,2014C has been low at all epochs. 
For the {JWST} observation, we already show above that the optical depth is only $\sim 0.15$ outside of the strong silicate band. 
For the {Spitzer} observations, we calculate the IR optical depth to be $\sim 0.5$ at 500 days, decreasing to 0.04 at 1920 days, in the observed bands.
Because of the low optical depth, the blackbody radius $r_{\rm bb}$ is much smaller than the dust location $r_{\rm dust}$ at all epochs, and it does not serve as a good indicator of the dust location. 
% In fact, the dust observed with Spitzer is most likely pre-exisiting, from it
The much larger $r_{\rm bb}$ at the {JWST} epoch is due entirely to the lower temperature.

We instead turn to the dust mass and temperature evolution to discern the dust origin. 
The dust temperature and mass evolution (Figures~\ref{fig:LTM_evo}a and \ref{fig:LTM_evo}b) are relatively flat in the range 500--2000 days, even at the epoch including observations around 10 $\mu$m. 
\cite{Tinyanont2019} conclude that the observed dust is pre-existing in the CSM, getting heated by the interaction front. 
This is because the CSM has a wind-like density profile $\rho \propto r^{-2}$ (inferred from light-curve fitting), and the outgoing shock has a surface area growing as $A \propto r^2$. 
Thus, the mass of CSM dust being heated by the shock is roughly constant, assuming that shock heats the CSM dust out to some distance in front of it.
% The resulting minimum dust mass required to fit the SED, about 0.005~$M_\odot$, is {roughly} consistent with the total CSM mass of a few $M_\odot$, {provided a typical gas-to-dust mass ratio of 100}. 
% The resulting minimum dust mass required to fit the SED in this phase, about 0.005~$M_\odot$, is {roughly} consistent with the total CSM mass of a few $M_\odot$, {provided a typical gas-to-dust mass ratio of 100}. 

We do caution that caveats applied for previous observations with minimal mid-IR coverage. 
At epochs between 1620 and 1900 days, there are only observations in the \textit{Spitzer} 3.6 and 4.5 $\mu$m bands, which are insensitive to further cooling, and the dust could be cooling and increasing during this time. 
The SED fit to the \textit{Spitzer} and ground-based mid-IR photometry from $\sim$1620 days presented in \citet{Tinyanont2019} is not unique, and cannot rule out a more massive dust component ($\sim$0.03 $M_\odot$) at $\sim$ 250 K. 
The fit simply used the least amount of dust to explain the observed SED. 
Using a more massive and cooler component to explain the 10~$\mu$m flux does not result in a better fit. 
In addition, we do not expect cold, newly formed dust, in addition to the observed $\sim$500 K dust at this phase. 
Ejecta dust would have created red-wing suppression of optical lines that we do not observe (Section~\ref{sec:spec}).
There could be new dust in the {CDS between the forward and reverse shocks} at this phase, but it should not be colder than what we later observed with \textit{JWST} as the interaction flux was higher at 1620 days.
% Further, such a large dust mass has never been observed in SNe of any type at such phase (Figure~\ref{fig:dust_mass_evo}). 
A detailed hydrodynamical simulation would be required to predict the dust temperature evolution.

% This evolution abruptly changes at the {JWST} epoch; the dust mass increases by an order of magnitude, while the temperature plummets by half.
The \textit{JWST} observations at 3477 day show that the minimum dust mass required to fit the spectrum has increased by an order of magnitude, while the temperature plummets by half.
% We emphasize that we have ground-based mid-IR imaging out to 12~$\mu$m at 1620 d (Figure~\ref{fig:spec_comp}), and they did not show dust mass and temperature that are inconsistent with {Spitzer}-only observations. 
As a sanity check, we perform dust fitting on synthetic photometry in the same photometric bands ($K_s$, $L$, $M$, \textit{Spitzer} 3.6 and 4.5 $\mu$m, and Subaru/COMICS N9.7, N10.5, and N11.7) using the \textit{JWST} spectrum, and also find a large increase in dust mass. 
Therefore, this mass increase is not due to the increased wavelength coverage of the \textit{JWST} data. 
% Prior to 1920 days, Spitzer two-band observations are still probing the same dust component.
% If this component is cooling, we should have observed some color evolution. 
We also know from the luminosity evolution (Figure~\ref{fig:LTM_evo}c) that the CSM profile is likely smooth between the {Spitzer} and {JWST} observations, so the increase in the observed dust mass cannot be explained by an additional dense CSM component.

The most likely scenario that explains this observation is that new dust has formed in the cold dense shell (CDS) between the forward and reverse shocks \citep{Smith2008, Sarangi2022}. 
At this epoch, there is now enough CSM ($\sim$1.5 $M_\odot$ from Section \ref{sec:bolo}) and SN ejecta (a few $M_\odot$, typical for SNe Ib; \citealp{Lyman2016}) processed by the forward and reverse shocks. %, respectively, into the CDS to condense the observed amount of dust. 
% The CDS now likely has $\sim 5 \ M_\odot$, sufficient to condense the observed amount of dust (0.078 $M_\odot$) at a plausible gas-to-dust ratio of $\sim$60.   
{The few $M_\odot$ of material in the CDS is sufficient to condense the observed amount of dust (0.078 $M_\odot$) at a plausible gas-to-dust mass ratio.}  
% The lower temperature of this newly formed dust is due to the lower ambient temperature in the CDS the better shielding from the interaction flux from the CDS's higher density. 
The newly formed dust has lower temperatures likely due to the higher density of the CDS shielding it from the interaction flux. 

Dust formation in the CDS is also consistent with the lack of red-wing absorption observed in the optical lines originating from the reverse shock (Section~\ref{sec:spec}). 
This is because the CDS dust would be totally outside of the line-emitting region, absorbing both redshifted and blueshifted parts of the line equally. 
Dust inside the ejecta, on the other hand, would preferentially absorb the redshifted part of emission lines formed outside, which is the case for all spectral lines seen in SN\,2014C at this epoch, as we discuss in the next section.

\begin{figure}
    \centering
    \includegraphics[width=\linewidth]{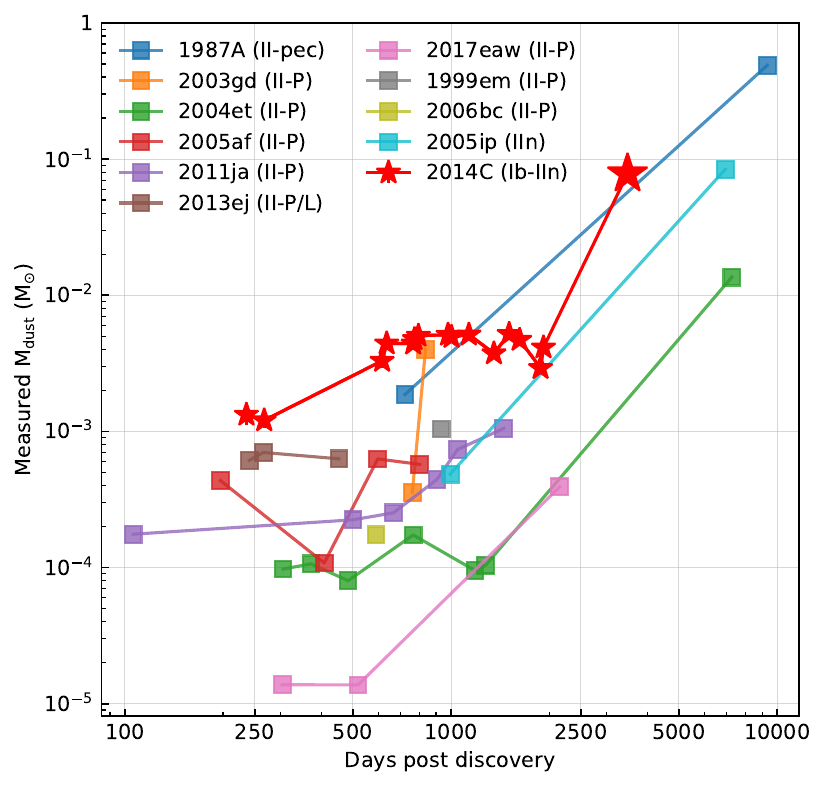}
    \caption{The evolution of the observed dust mass in SN\,2014C compared with select CCSNe. 
    These comparison objects are from \citet{Shahbandeh2024} and references therein.
    }
    \label{fig:m_evo_comp}
\end{figure}

Figure~\ref{fig:m_evo_comp} compares the observed dust mass evolution of SN\,2014C to those in the literature, including the recent late-time observations of SNe II-P 2004et and 2017eaw \citep{Shahbandeh2023} and SN\,2005ip \citep{Shahbandeh2024}.
The plot shows that the dust mass in SN\,2014C evolves in the same manner as that in other CCSNe. 
We note that, until recently, there were few mid-IR measurements for SNe at early time, and the cold dust mass, missed by these observations, could be significantly higher than what was observed. 
A sample of CCSNe with \textit{JWST} observations throughout their evolution is crucial to determine whether this is the case.

% The lack of red-wing absorption suggest minimal dust in the SN ejecta. 

% This new dust formation does not impart strong red-wing absorption in the optical lines possibly because of the disk-like CSM geometry. % inferred by \citet{Thomas2022}.
% Such an absorption only happens if the dust is totally interior to the line forming region, absorbing light from the receding ejecta.
% This lack of red-wing absorption in the optical lines also rules out dust formation in the ejecta. 
% \citet{Thomas2022} inferred from spectroscopic observations along with X-rays that there must be both high and low density regions around SN~2014C, to allow for both low absorption X-ray emission and strong ongoing interaction lines. 
% Different lines also originate from different regions, based on their FWHM.
% H$\alpha$ comes from above the forward shock in the boundary layer between the CSM disk and the free-flowing SN ejecta above and below it; and broader forbidden lines come from the reverse shock. 
% In this scenario, if the new dust is condensing in the CDS (or rather a cold dense ring here), it will not preferentially absorb the red wing of these lines, as long as we are not observing the system edge on.
% We validate these observations with our own analysis of optical through mid-IR spectra in the next Section.

\section{Spectral Lines Evolution}\label{sec:spec}
% Figure~\ref{fig:spec_lines} show selected notable spectroscopic lines in SN\,2014C. 
Optical to IR spectra of SN\,2014C continue to be rich in features at the most current epoch, more than 10~yr post-explosion (Figure~\ref{fig:new_spec}). 
In addition to our new observations at 3121, 3230, and 3789 day post-explosion, we analyze the Keck/DEIMOS spectrum from 269 days and Keck/LRIS spectrum from 617 days published by \citet{Mauerhan2018}, as well as optical spectra presented by \citet{Thomas2022}.
The latest optical spectrum at 3789 days is dominated by strong H$\alpha$ along with forbidden oxygen lines.
{High-ionization iron lines such as [\ion{Fe}{7}] $\lambda$6087 and [\ion{Fe}{11}] $\lambda$7892 disappear before our new observations, as expected by weakening CSM interactions.}
%with a FWHM of \textbf{xxxx} $\rm km\, s^{-1}$, which is narrower than other lines as noticed by \citet{Thomas2022}. 
The last ground-based NIR spectrum shows a clear detection of only the \ion{He}{1} 1.083 $\mu$m line (shown in Figure~\ref{fig:forbiddenO}). 
Finally, the MIRI MRS spectrum exhibits a strong [\ion{Ne}{2}] 12.813 $\mu$m line. 
Owing to the strong galaxy background, reliable flux calibration of the optical lines is not possible given the lack of reliable photometry.
The only line for which we have absolute flux calibration is [\ion{Ne}{2}] 12.813 $\mu$m. 
These emission lines demonstrate different profiles, indicative of the geometry of their respective formation regions. 

\subsection{H$\alpha$ from the CSM}
The only hydrogen line left detectable with an intermediate-width ($\sim$1000 $\rm km\,s^{-1}$) component at late times is H$\alpha$.
The progenitor star of SN\,2014C is hydrogen poor, so all hydrogen emission comes from the CSM. 
% This explains why the FWHM observed in hydrogen lines is much lower than that of other lines as the CSM never gets accelerated to the ejecta velocity. 
{We fit narrow and broad components to the H$\alpha$ line simultaneously with the [\ion{N}{2}] $\lambda$$\lambda$6548, 6583 from the host.
}
As also measured by \citet{Thomas2022}, the FWHM of the broad H$\alpha$ is $\sim 1000$ $\rm km\, s^{-1}$ at all epochs. 
This is much lower than the FWHM observed in other lines, as the CSM never gets accelerated to the ejecta velocity.
We interpret this as the velocity of the shocked CSM in the CDS between the forward and reverse shocks. 

\subsection{He~I 1.083 and 2.059 $\mu$m Lines}
The He~I 1.083 and 2.059 $\mu$m lines have been observed at 1354 and 1693 days post-explosion, and only the 2.059 $\mu$m line at 275 days, at  high S/N by \citet{Tinyanont2019}.
In the new observation at 3807 days, only the stronger 1.083 $\mu$m line is clearly detected.  
Figure~\ref{fig:forbiddenO} (top) shows these lines, in addition to \ion{He}{1} $\lambda$5876 in comparison with [\ion{O}{3}].
We use this oxygen line because it is the only line that persists throughout all phases for which we have NIR observations.
% At the 275 day, we also plot the \ion{He}{1} 5876 \AA\ line from a spectrum from \citet{Mauerhan2018}.
% This line fades in the subsequent epochs shown here. 

In Figure~\ref{fig:forbiddenO} (middle), we compare line profiles of the NIR helium lines.
The line profiles of He~I 1.083 and 2.059 $\mu$m are broadly similar over the first two epochs; they can be fit with one broad Gaussian (FWHM $\approx$ 4500 $\rm km\, s^{-1}$) and two intermediate-width Gaussians (FWHM $\approx$ 1700 $\rm km\, s^{-1}$) at around $-4000$ and 0 $\rm km\, s^{-1}$ \citep{Tinyanont2019}, in addition to unresolved narrow components likely from the host galaxy.\footnote{\citet{Tinyanont2019} reported the standard deviation of the Gaussian, but erroneously marked it as FWHM.} 
To be consistent with \citet{Tinyanont2019}, we refer to these components as ``a,'' ``b,'' and ``c,'' respectively.
% The blueshifted ``b'' component is present in both the 1.083 and 2.059 $\mu$m lines, so it is not due to a different line. 

% The velocity evolution of these component is notable.
Figure~\ref{fig:forbiddenO} (middle) shows Gaussian fits to the He~I 1.083 $\mu$m line.
The broad component ``a'' has a mean velocity close to zero at all epochs. 
The FWHM evolves from 8700 to 5900 $\rm km\, s^{-1}$.
The velocity and its evolution are consistent with this component arising from the SN ejecta heated by the reverse shock.
The decreasing velocity corresponds to the reverse shock propagating deeper into the lower-velocity part of the ejecta. 

% At 3807 day, the line profile could also be interpreted as a broader Gaussian getting absorbed in the blueshifted part, around $-1800 \ \rm km\,s^{-1}$.
% This would be expected if the reverse shock has passed through the helium-rich part of the ejecta, and is now illuminating the approaching part of it from behind. 
% This could explain the line profile observed in Figure~\ref{fig:forbiddenO} top. 

At the first epoch, the centered intermediate-width ``c'' component is significantly redshifted by 470 and 840 $\rm km\, s^{-1}$ in the He~I 1.083 and 2.059 $\mu$m lines, respectively. 
The similar redshifted feature is seen in the [\ion{O}{3}] line at the same epoch as well (Figure~\ref{fig:forbiddenO}, top; data from \citealp{Thomas2022}). 
The ``c'' component of the 1.083 $\mu$m line is close to $v = 0$ by 1695 days, while the 2.059 $\mu$m line is still significantly redshifted. 
% The FWHM of this component is $\sim$4 smaller than that of the broad component, which corresponds to the jump condition of a non-radiative shock.
% We speculate that this is due to the 2.059 $\mu$m line
As such, this component is likely from the ejecta that pass through the reverse shock into the CDS.

Lastly, there is a blueshifted intermediate-width component ``b'' centered at around $-4000 \ \rm km\, s^{-1}$.
This component is intrinsic to the helium lines, as it is present in both He~I 1.083 and 2.059~$\mu$m.
Like the ``c'' component, its FWHM remains relatively constant at 1200~$\rm km\, s^{-1}$.
The velocity of the line center of this blueshifted component evolves from $-4100$ to $-3800 \ \rm km\, s^{-1}$ over the first two epochs.

In the last epoch at day 3807, this component turned into absorption (see Figure~\ref{fig:forbiddenO}, middle right), and the line profile can be interpreted as a broad Gaussian getting absorbed in the blueshifted part, around $-1800 \ \rm km\,s^{-1}$.
% This indicates that the reverse shock has passed through the helium-rich part of the ejecta, and is now illuminating the part approaching the observer from behind. 
This indicates that the reverse shock has passed through the helium-rich part of the asymmetric ejecta responsible for the blueshifted component, and is now illuminating the part approaching the observer from behind. 
% This could explain the line profile observed in Figure~\ref{fig:forbiddenO} top. 

The lack of a companion redshifted component could be due to dust obscuring the receding part of the ejecta responsible for component ``b''; however, we argue that this is not the case. 
As discussed earlier in Section~\ref{sec:dust}, the dust in SN\,2014C has grown in mass by an order of magnitude between 2000 and 3500 days post-explosion, but the helium line profile has not evolved significantly. 
% Other components can be fit with symmetric Gaussian profiles. 
Lastly, [\ion{Ne}{2}] 12.813 $\mu$m, which we will later discuss, shows a profile very similar to that of the optical lines, despite dust having very small opacity at that wavelength compared with at 1 $\mu$m.

% At 3807 day, the He 1.083 $\mu$m line still has a similar profile.
% The broad component is with FWHM$=$2700 $\rm km\, s^{-1}$; the intermediate-width components remain at FWHM $\sim$ 1700 $\rm km\, s^{-1}$, but the velocity of the blueshifted component is at \textbf{XXX?} $\rm km\, s^{-1}$. 

\subsection{Forbidden Oxygen Lines}
Forbidden oxygen lines are prominent in the latest optical spectrum at 3789 days; 
we identify the [\ion{O}{1}] $\lambda\lambda$6300, 6364, [\ion{O}{2}] $\lambda\lambda$7319, 7330, and [\ion{O}{3}] $\lambda\lambda$4959, 5007 doublets.
The feature at $\sim 7319$ \AA\ was identified as [\ion{Ca}{2}] $\lambda\lambda$7291, 7324 by \citet{Thomas2022}.
We disfavor this identification because the two relatively widely spaced [\ion{Ca}{2}] doublets are equally strong and should lead to a broader line profile. 
The profile we observe for this feature is more similar to that of [\ion{O}{1}] and [\ion{O}{3}], when centered around the stronger [\ion{O}{2}] $\lambda$7319 line. 
Further, the two lines that are more likely [\ion{Ca}{2}] are present in the spectrum at 269 days, along with the \ion{Ca}{2} NIR triplet. 
These features faded by 617 days.
% both [\ion{O}{1}] and [\ion{O}{3}] exist, we favor the [\ion{O}{2}] identification.
Indeed, similar arguments were made in favor of [\ion{O}{2}] over [\ion{Ca}{2}] in late-time spectra of SN\,1980K \citep{Fesen1999}. 
Figure~\ref{fig:forbiddenO} (bottom) compares the line profiles of the  forbidden oxygen lines at three epochs post-explosion. 

The [\ion{O}{1}] and [\ion{O}{3}] doublets show a strong, narrow component in the 617 day spectrum from the unshocked CSM heated by the interaction flux (this component is much brighter than the host emission at this epoch). 
It gets weaker with time relative to the underlying broad component.
% While more difficult to fit compared with the helium lines due to their doublet nature, we estimate the FWHM of these lines to be \textbf{YYY}.
We agree with the interpretation of \citet{Thomas2022} that the broad component of forbidden lines arises from the SN ejecta heated by the reverse shock based on their higher velocity compared with the H$\alpha$ line, and the velocity similar to that of the broad helium lines. 

In addition to the forbidden oxygen lines, we note that [\ion{S}{3}] $\lambda$9531 is detected between 3121 and 3789 days. 
Its line profile is similar to that of oxygen.
This line was outside of the spectral coverage in the \citet{Thomas2022} dataset.

Most notably, the forbidden oxygen lines all show a blueshifted component at around $-3000 \ \rm km\, s^{-1}$, very similar to what is observed in the helium lines. 
At 1300 and 1650 days, the blueshifted component in the oxygen line is still not obvious (Figure~\ref{fig:forbiddenO}, top), but this feature becomes strong by day 3121  (Figure~\ref{fig:forbiddenO}, bottom). 
Because these forbidden lines are optically thin, their profile reflects the gas distribution.
The similarity in the line profile across multiple species at different wavelengths suggests that this feature is due to the asymmetry in the common line-forming region, likely the CDS, for these intermediate-width lines. 

\subsection{[Ne II] 12.813 $\mu$m and [Ne III] 15.550 $\mu$m}
\begin{figure}
    \centering
    \includegraphics[width=0.8\linewidth]{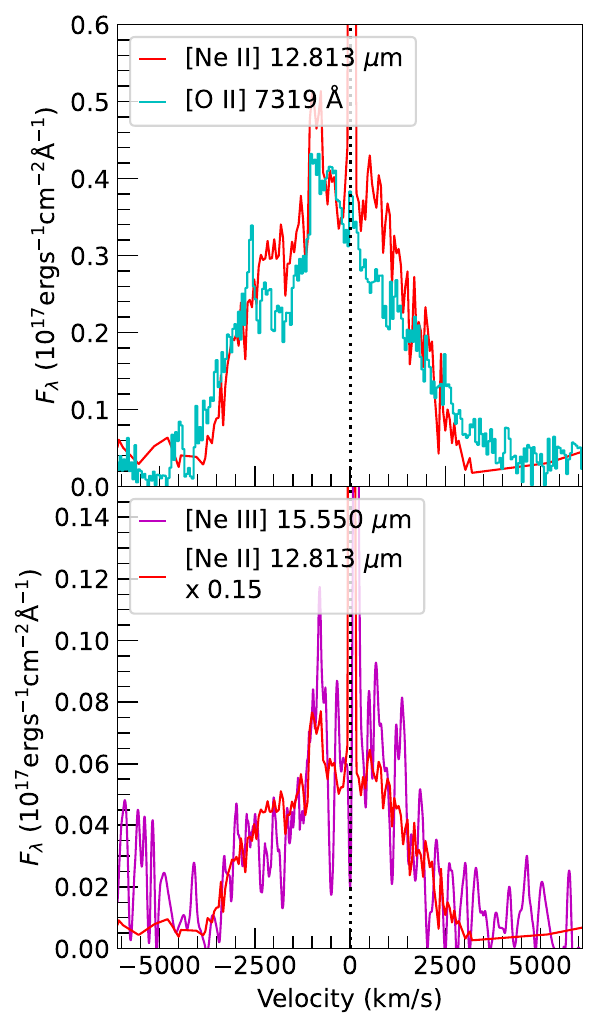}
    \caption{MIRI MRS spectrum showing [\ion{Ne}{2}] 12.813 $\mu$m (top) and [\ion{Ne}{3}] 15.550 $\mu$m (bottom).
    The absolute flux is provided in $F_\lambda$ and the dust continuum has been subtracted. 
    In the top panel, the scaled line profile of [\ion{O}{2}] $\lambda$7319 from day 3789 is provided for comparison.
    In the bottom panel, [\ion{Ne}{2}] 12.813 $\mu$m scaled by a factor of 0.15 is shown for comparison. 
    }
    \label{fig:Ne_MIR}
\end{figure}

The only strongly detected line from the {JWST}/MRS data that has a broad component is [\ion{Ne}{2}] 12.813 $\mu$m.
The [\ion{Ne}{3} 15.550 $\mu$m line is marginally detected. 
Figure~\ref{fig:Ne_MIR} shows these two lines in velocity space.

Figure~\ref{fig:14C_image} (top) demonstrates that there are several star-forming knots in the host galaxy that are luminous in [\ion{Ne}{2}] 12.813 $\mu$m.
However, the flux presented here, especially the broad component, is from the SN. 
{We note that the broad component seen under the [\ion{Ne}{2}] line in the background spectrum in Figure~\ref{fig:14C_image} (bottom) is from the underlying PAH feature at 12.7 $\mu$m, and its subtraction residual does not significantly affect the broad component of the [\ion{Ne}{2}] 12.813 $\mu$m line from the SN.}
This line has recently gained attention as the primary coolant for systems with magnetar heating \citep{Dessart2024}. 
Forbidden oxygen lines are also a significant coolant in those systems, and have been used as evidence for a pulsar wind nebula in SN\,2012au \citep{Milisavljevic2018}.
The presence of these lines in SN\,2014C, along with H$\alpha$, suggests that CSM interaction can also excite these forbidden lines, complicating the interpretation.  
In addition, we note that forbidden oxygen and mid-IR neon (and other forbidden) lines are also present in SN remnants \citep[e.g.,][]{Kravtsov2024, Milisavljevic2024}.

% \citet{Kravtsov2024} also recently shown that these 
% making interpretation complicated. 

The [\ion{Ne}{2}] line in SN\,2014C has a generally similar profile to that of the forbidden oxygen lines.
Because the dust optical depths at 0.7 and 12.8 $\mu$m differ by 2 orders of magnitude, the line profile must be intrinsic to the geometric distribution of gas and not due to preferential dust absorption. 
The profile can be fitted with a narrow, unresolved, Gaussian component at rest and a broad Gaussian (FWHM $=4500 \ \rm km\,s^{-1}$) centered at $-440 \ \rm km\, s^{-1}$.
The blueshifted intermediate-width component is less pronounced in this line. 

With absolute flux calibration, we can use this line to constrain the location of the emission region. 
We integrate the line luminosity of the broad component, excluding the narrow component that could be from host contamination. 
The line luminosity in the broad component is $L_{\rm [Ne\ II]} = (2.4 \pm 0.1)\times 10^{38} \ \rm erg \, s^{-1}$. 
For an optically thin line, the luminosity is a volume integral
\begin{equation}
    L = \int 4 \pi j_{ki} \,  dV = \int n_k A_{ki} h \nu_{ki} \, dV,
\end{equation}
where $j_{ki}$ is the emission coefficient of this transition, $n_k$ is the number density of the ions in the upper state, $A_{ki}$ is the Einstein A coefficient, $h$ is the Planck constant, and $\nu_{ki}$ is the frequency of this line.  
If we assume a constant-density shell of emission with inner and outer radii $R_i$ and $R_o$, we get 
\begin{equation}
    L = \frac{4}{3} \pi n_k A_{ki} h \nu_{ki} (R_o^3 - R_i^3).
\end{equation}
The Einstein A coefficient of this line is $A_{21} = 8.59\times10^{-3} \rm \ s^{-1}$.
The density range where we expect to see forbidden lines is $10^3$ to $10^5 \ \rm cm^{-3}$.
With the lower-density end at $n = 10^3 \rm \ cm^{-3}$, $(R_o^3 - R_i^3) = 4.3\times 10^{49} \ \rm cm^3$.
If we assume that the emission region is a filled sphere with $R_i = 0$, the radius is $R_o = 3.5 \times 10^{16} \rm \ cm$. 
Coincidentally, this is roughly the same as the radius of the CSM-free inner bubble determined from the time delay to the onset of interaction \citep{Margutti2017}.
If we assume that the emission region is a shell with $R_i >0$, the shell becomes thin very quickly. 
With $R_i = 2.6 \times 10^{16} \ \rm cm$, the fractional thickness of the shell is already $(R_o - R_i)/R_i = 0.5$. 
Such thin shells would produce a flat-topped or double-peaked line profile, inconsistent with our observations. 
As such, our data suggest that the [\ion{Ne}{2}] 12.813 $\mu$m line originates primarily from the SN ejecta, heated at this epoch by the offset CSM, resulting in the overall blueshift of the broad component. 
The lack of an intermediate component  indicates that the reverse shock has not significantly processed the Ne-rich part of the ejecta yet. 

Lastly, we note that we do not detect other mid-IR lines predicted by \citet{Dessart2024} with an intermediate-width component. 
Specifically, we checked the [\ion{Ni}{2}] 6.634,  [\ion{Ar}{2}] 6.983, [\ion{Ni}{3}] 7.347, and [\ion{Ar}{3}] 8.989, and [\ion{S}{3}] 18.708~$\mu$m lines, and only detect narrow emission from the host in most cases.  

\subsection{Ejecta and CSM Geometry Constrained by Emission Lines}
The consistent profile across many optically thin forbidden lines and the helium lines point to the complicated CSM geometry of SN\,2014C. 
Given the size constraint and the relatively symmetric line profile, we argue that [\ion{Ne}{2}] 12.813 $\mu$m originates from the inner ejecta, heated by emission from the reverse shock. 
Because of the similarity in the line profile, the broad component of helium and forbidden oxygen lines likely come from the inner ejecta.
The decreasing FHWM velocity of this component observed most clearly in the helium lines supports this interpretation, as the reverse shock traverses to the inner ejecta with lower velocity. 

\begin{figure*}
    \centering
    \includegraphics[width=1\linewidth]{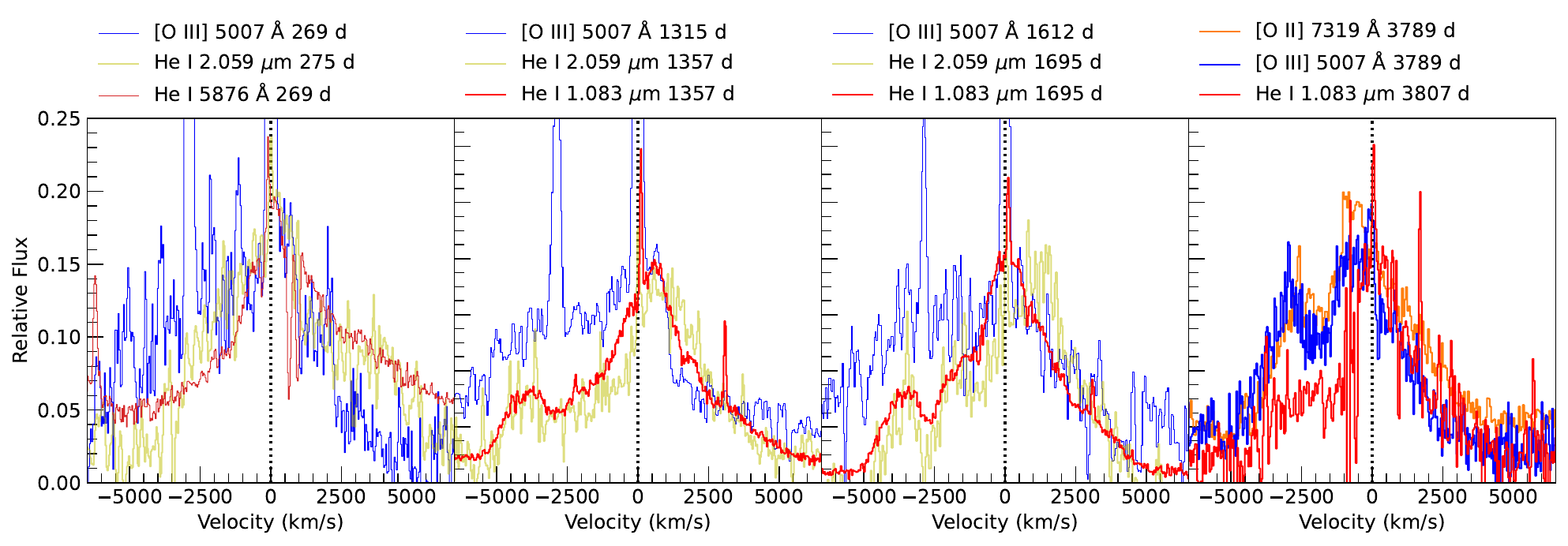}
    \includegraphics[width=1\linewidth]{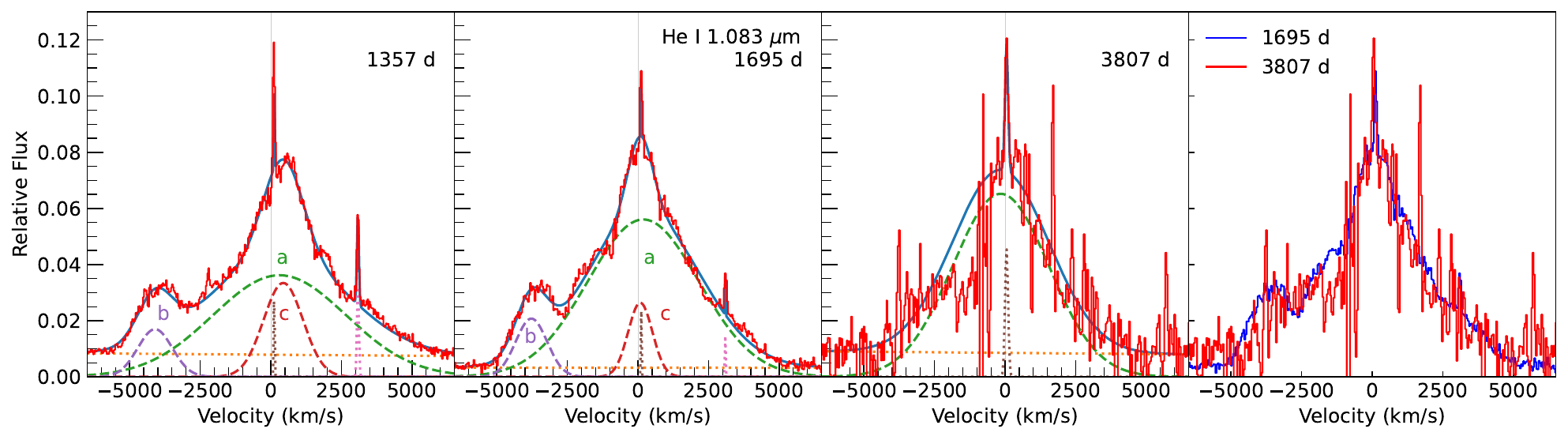}
    \includegraphics[width=1\linewidth]{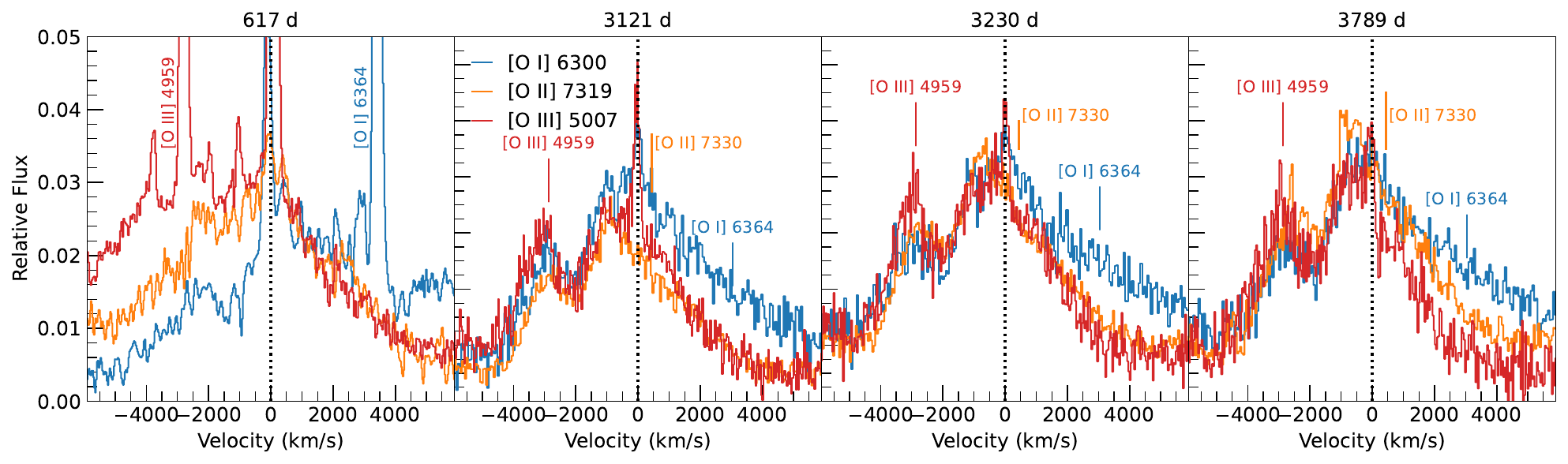}
    \caption{\textbf{Top:} comparisons between the He~I 5876~\AA, 1.083 $\mu$m, and 2.059 $\mu$m lines, as well as [\ion{O}{3}] 5007~\AA\ at around 270, 1300, 1650, and 3800 days post-explosion. The optical spectrum at 269 d showing \ion{He}{1} $\lambda$5876 and [\ion{O}{3}] is from \citet{Mauerhan2018}. The optical spectra at 1315 and 1612 days showing [\ion{O}{3}] are from \citet{Thomas2022}. The NIR spectra from 275, 1357, and 1695 days showing helium lines are from \citet{Tinyanont2019}. 
    \textbf{Middle:} model fit to the  \ion{He}{1} 1.083 $\mu$m line profile shown in the top panel. The 1357 and 1695 day epochs are presented by \citet{Tinyanont2019}. They can be fit with a broad component (``a'') likely from the ejecta, and two intermediate-width components likely from the CDS. 
    Component ``b'' is blueshifted to about $-4000 \ \rm km\,s^{-1}$, while ``c'' is at rest. 
    The 3807 d epoch can be fitted with one broad Gaussian with some flux absorbed around $-1800 \ \rm km\,s^{-1}$. 
    The rightmost panel compares the profile of the 1.083 $\mu$m line at 1695 and 3807 days to highlight the potential switch of the blueshifted component from emission to absorption.
    \textbf{Bottom:} comparisons between forbidden oxygen doublets [\ion{O}{1}], [\ion{O}{2}], and [\ion{O}{3}] at 617, 3121, 3230, and 3789 days. The first-epoch spectrum is from \citet{Mauerhan2018}.
    The spectral profiles of the three lines are similar in the last two epochs, and only evolve slightly between those epochs.
    % The forbidden oxygen lines and the helium lines share
    }
    \label{fig:forbiddenO}
\end{figure*}

The mean velocity evolution observed in several lines points to a CSM distribution in which its center is slightly offset from the SN location. 
First, at 1357 days, the broad and intermediate-width (from the CDS) components of the helium lines are redshifted by around 400 $\rm km\,s^{-1}$.
A similar shift is also seen in the [\ion{O}{3}] line at a comparable epoch (Figure~\ref{fig:forbiddenO}, top). 
This could be explained if the CSM is closer to the SN on the side away from us; 
the SN shock interacts with that part of the CSM more strongly at early times, producing redshifted helium lines and the ledge feature seen in [\ion{O}{3}] at similar epochs.
We note that, while likely originating from a merger \citep[e.g.,][]{Morris2007}, the equatorial CSM ring around SN\,1987A is also offset from the SN, with the shock interaction starting on the northeast portion of the ring in 1995 \citep{Sonneborn1998, Lawrence2000}.
\citet{Sonneborn1998} also reported  blueshifted H$\alpha$ emission without an associated redshifted component, which they attributed to an interaction with an inward protrusion of the CSM ring. 

The CSM on the side away from the observer is overcome sooner, and by 3121 days, the interaction is dominated by the CSM on the side closer to us.
This leads to the significant mean velocity of $-440 \ \rm km\,s^{-1}$ observed in the [\ion{Ne}{2}] line, which suggests that the reverse shock (which is the power source) is stronger in the part of the ejecta approaching the observer at this phase.
% Determining the mean velocity in forbidden oxygen lines is difficult because they are all doublets. 
The enhanced CSM in the side closer to us also leads to the blueshifted intermediate-width component observed in the forbidden oxygen lines at around $-3000 \ \rm km \, s^{-1}$, and the blueshifted broad component of [\ion{Ne}{2}]. 
At this epoch, the reverse shock has likely passed through the helium-rich part of the ejecta.
As a result, the blueshifted helium emission appears in absorption by 3807 days post-explosion (Figure~\ref{fig:forbiddenO}, middle). 
% If this picture is correct, as the reverse shock passed through the oxygen-rich ejecta into the neon-rich part of the ejecta, we should see the blueshifted intermediate width component of the [\ion{Ne}{2}] line appear and the blueshifted intermediate width component of the forbidden oxygen lines turning into absorption.

The prolonged presence of the blueshifted intermediate-width component in optically thick lines points to an asymmetric distribution of material confined to a small range of ejecta velocity in the SN. 
In the long-lasting SN IIn KISS15s, \citet{Kokubo2019} reported a similar line profile in H$\alpha$, and they associated it with a plume of high-velocity ejecta expanding through a low-density region in the CSM. 
A similar picture could be the case for SN\,2014C, but with hydrogen-poor, helium-rich ejecta, so this feature only appears in helium and forbidden metal lines, not hydrogen.
If this is the case, then the CSM has to take up more of the solid angle around the SN than a disk previously proposed in the literature.
Because there is no redshifted component even when the dust optical depth is low, it is unlikely that there is an opposite plume hidden from view.
% This indicates further asymmetry in the CSM around SN\,2014C. 

\section{Conclusion}\label{sec:conclusion}
SN\,2014C remains an explosion in action for more than a decade. 
Optical and IR spectra still show emission lines excited by the ongoing interactions. 
Persistent H$\alpha$ at $\sim$1000 $\rm km\ s^{-1}$ demonstrates that the interaction with hydrogen-rich CSM is still ongoing, and this line emerges from the shocked CSM. 
Forbidden oxygen lines in the optical, helium lines in the NIR, and a forbidden neon line in the mid-IR reveal a complex geometry of the CSM. 
They all have a broader-width component $\sim 5000 \ \rm km\, s^{-1}$, likely excited by the reverse shock propagating back into the ejecta.
They also show an intermediate-width $\sim1000 \ \rm km\,s^{-1}$ component from the slower CDS. 
The helium and oxygen lines exhibit an additional blueshifted intermediate-width component, centered at around $-4000 \ \rm km\,s^{-1}$, that could be from an ejecta plume pointing toward the observer. 
This component of the helium line turns from emission into absorption, which is expected in this scenario as the reverse shock traverses the helium-rich part of the ejecta.

Our {JWST} NIRSpec and MIRI observations capture the bulk of the bolometric luminosity, and show that SN~2014C is still following a decline predicted by a semianalytic model of CSM interaction fit to the {Spitzer} data by \citet{Tinyanont2019}. 
From that model, the CSM is wind-like with $\rho \propto r^{-2}$, but need not be isotropic.  
The shock is at a radius of $\sim 3\times 10^{17} \ \rm cm$ at the time of the {JWST} observations, and has interacted with at least 1.6 $M_\odot$ of CSM.

Fitting the mid-IR continuum, we find that 0.0780 $M_\odot$ of dust at $\sim 245$\,K has formed, likely in the CDS between the forward and reverse shocks. 
This dust mass increases by a factor of $\sim$10 from the last observations around days 1600--2200, for which we have mid-IR imaging from the ground. 
A dust optical depth analysis shows that the dust is likely optically thin, and the mass observed reflects the true mass of the dust at temperatures detectable with MIRI. 
This is among the highest dust mass in an SN observed around this epoch.
Along with recent {JWST} observations of SNe~IIn (e.g., SN\,2005ip, \citealp{Shahbandeh2024}), and upcoming results from the GO-1860 (PI Fox) program, we find that interacting SNe produce significantly more dust than noninteracting objects.
The relative rate of interacting SNe in the early Universe must be better modeled to accurately account for the dust observed in nascent galaxies.

\bigskip
\bigskip

% \begin{acknowledgments}
Part of this research benefited from the {JWST} Data Analysis and Processing Workshop (South East Asia), organized by the National Astronomical Research Institute of Thailand (NARIT) in July 2024 with support from the IAU Hands-On-Workshops and COSPAR Capacity Building programs. 
S.T. thanks N.\ Leethochawalit and K.\ Chanchaiworawit for leading the organization effort, and J.\ Alvarez Marquez for his instruction on NIRSpec IFU and MIRI MRS data reduction.
He also thanks T. Moriya and C. Fransson for useful discussions. 

This work is based in part on observations made with the NASA/ESA/CSA {James Webb Space Telescope}. The data were obtained from the Mikulski Archive for Space Telescopes at the Space Telescope Science Institute, which is operated by the Association of Universities for Research in Astronomy, Inc., under NASA contract NAS 5-03127 for {JWST}. These observations are associated with program GO-2348.
This work was supported by National Aeronautics and Space Administration (NASA) Keck PI Data Awards, administered by the NASA Exoplanet Science Institute. 
S.T. and K.W. acknowledge support by the Fundamental Fund of Thailand Science Research and Innovation (TSRI) grant FFB680072/0269 through NARIT. 
K.M. acknowledges support from JSPS KAKENHI grants JP24H01810 and 24KK0070.
The UCSC team is supported in part by STScI grants JWST-GO-2348, JWST-DD-6659, and JWST-DD-6838; NASA grants 80NSSC23K0301 and 80NSSC24K1411; and a fellowship from the David and Lucile Packard Foundation to R.J.F.
Financial support to L.G.\ is acknowledged from the Spanish Ministerio de Ciencia, Innovaci\'on y Universidades (MCIU) and the Agencia Estatal de Investigaci\'on (AEI) 10.13039/501100011033 under the PID2020-115253GA-I00 HOSTFLOWS and PID2023-151307NB-I00 SNNEXT projects, from Centro Superior de Investigaciones Cient\'ificas (CSIC) under the projects PIE 20215AT016, ILINK23001, COOPB2304, the program Unidad de Excelencia Mar\'ia de Maeztu CEX2020-001058-M, and from the Departament de Recerca i Universitats de la Generalitat de Catalunya through the 2021-SGR-01270 grant.
A.V.F.'s group at UC Berkeley has received financial assistance from the Christopher R. Redlich Fund, Gary and Cynthia Bengier, Clark and Sharon Winslow, Alan Eustace (W.Z. is a Bengier-Winslow-Eustace Specialist in Astronomy), William Draper, Timothy and Melissa Draper, Briggs and Kathleen Wood, Sanford Robertson (T.G.B. is a Draper-Wood-Robertson Specialist in Astronomy; Y.Y. was a Bengier-Winslow-Robertson Fellow in Astronomy), and numerous other donors.
Z.G.L.\ is supported by the Marsden Fund administered by the Royal Society of New Zealand, Te Apārangi under grant M1255.
C.L.\ acknowledges support under DOE award DE-SC0010008 to Rutgers University.
D.M.\ acknowledges support from the National Science Foundation (NSF) through grants PHY-2209451 and AST-2206532.
J.R.\ acknowledges financial support from a NASA ADAP grant (80NSSC23K0749).
T.T.\ acknowledges support from the NSF grant AST-2205314 and NASA ADAP award 80NSSC23K1130.
Q.W.\ is supported by the Sagol Weizmann-MIT Bridge Program.

Part of this work uses the Chalawan High-Performance Computer Cluster at NARIT.
Some of the data presented in this paper were obtained from the Mikulski Archive for Space Telescopes (MAST). STScI is operated by the Association of Universities for Research in Astronomy, Inc., under NASA contract NAS5-26555. Support for MAST for non-{HST} data is provided by the NASA Office of Space Science via grant NNX13AC07G and by other grants and contracts. 
Some of the data presented herein were obtained at Keck Observatory, which is a private 501(c)3 nonprofit organization operated as a scientific partnership among the California Institute of Technology, the University of California, and NASA. The Observatory was made possible by the generous financial support of the W. M. Keck Foundation. The authors wish to recognize and acknowledge the very significant cultural role and reverence that the summit of Maunakea has always had within the Native Hawaiian community. We are most fortunate to have the opportunity to conduct observations from this mountain.
This research has made use of the Keck Observatory Archive (KOA), which is operated by the W. M. Keck Observatory and the NASA Exoplanet Science Institute (NExScI), under contract with NASA.

%\end{acknowledgments}

\bibliography{2014C.bib}{}
\bibliographystyle{aasjournal}

\appendix
\setcounter{figure}{0}                       % <---------------
\renewcommand\thefigure{A.\arabic{figure}}

\section{Markov Chain Monte Carlo Sampling of the Dust Parameters}

\begin{figure*}[ht!]
    \centering
    \includegraphics[width=0.8\linewidth]{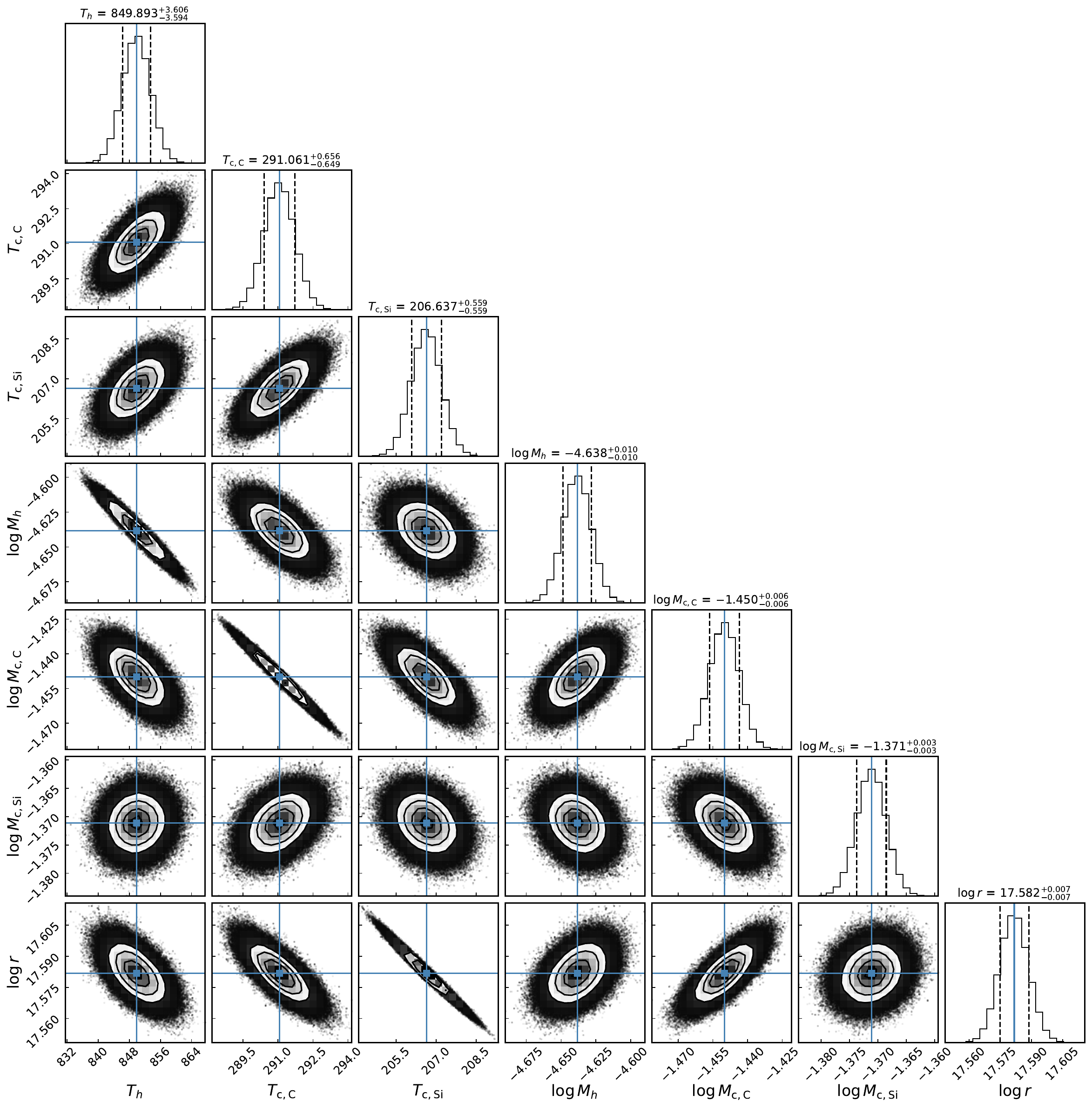}
    \caption{Corner plot of the posterior distribution of dust-fitting parameters used in modeling the IR spectrum of SN\,2014C.}
    \label{fig:corner}
\end{figure*}

We use the \texttt{emcee}  package \citep{emcee} to perform Markov Chain Monte Carlo sampling of the dust parameters presented in Section~\ref{sec:dust}. 
The free parameters in the fit are the temperature and mass of two carbonaceous dust components and one silicate dust component. 
The prior distributions are uniform with the allowed ranges $1 \leq T \leq 3000 \ \rm K$ and $10^{-7} \leq M \leq 2 \ M_{\odot}$. 
The log likelihood function is $\ln{\lambda} = -0.5 \sum_\lambda (y_\lambda - m_\lambda)^2 / (\sigma_\lambda^2)$, where $y_\lambda$ and $\sigma_\lambda$ are (respectively) the observed flux and uncertainty at wavelength $\lambda$, and $m_n$ is the model flux at that wavelength.
We run 32 walkers in 50,000 steps, and use the \texttt{get\_autocorr\_time} function of \texttt{emcee} to automatically determine the burn-in steps to remove.
The chains converge quickly after around 100 steps, and we remove the first 500 steps from the analysis. 
Figure~\ref{fig:corner} shows the corner plot of the posterior distribution of the fitted parameters.

\end{document}